\documentclass[%
 reprint,
superscriptaddress,
 amsmath,amssymb,
 aps,
]{revtex4-2}

\usepackage{graphicx}% Include figure files
\usepackage{bm}% bold math
\usepackage{placeins}
\usepackage{float}
\usepackage{svg}
\usepackage{afterpage}

\begin{document}

\preprint{APS/123-QED}

\title{Integration of through-sapphire substrate machining with superconducting quantum processors}

\author{Narendra Acharya}
\affiliation{%
 Oxford Quantum Circuits, Thames Valley Science Park, Shinfield, Reading, United Kingdom, RG2 9LH}%
  \author{Robert Armstrong}
\affiliation{%
 Oxford Quantum Circuits, Thames Valley Science Park, Shinfield, Reading, United Kingdom, RG2 9LH}%
\author{Yashwanth Balaji}
\affiliation{%
 Oxford Quantum Circuits, Thames Valley Science Park, Shinfield, Reading, United Kingdom, RG2 9LH}%
 \author{Kevin G.~Crawford}
\affiliation{%
 Oxford Quantum Circuits, Thames Valley Science Park, Shinfield, Reading, United Kingdom, RG2 9LH}%
  \author{James C.~Gates}
\affiliation{%
Optoelectronics Research Centre, University of Southampton, Southampton
SO17 1BJ, UK}%
 \author{Paul C.~Gow}
\affiliation{%
Optoelectronics Research Centre, University of Southampton, Southampton
SO17 1BJ, UK}%
 \author{Oscar W.~Kennedy}
\affiliation{%
 Oxford Quantum Circuits, Thames Valley Science Park, Shinfield, Reading, United Kingdom, RG2 9LH}%
  \author{Renuka Devi Pothuraju}
\affiliation{%
 Oxford Quantum Circuits, Thames Valley Science Park, Shinfield, Reading, United Kingdom, RG2 9LH}%
 \author{Kowsar Shahbazi}
\affiliation{%
 Oxford Quantum Circuits, Thames Valley Science Park, Shinfield, Reading, United Kingdom, RG2 9LH}%

\author{Connor~D.~Shelly}\email{cshelly@oqc.tech}
\affiliation{%
 Oxford Quantum Circuits, Thames Valley Science Park, Shinfield, Reading, United Kingdom, RG2 9LH}%

\date{\today}% It is always \today, today,

\begin{abstract}
We demonstrate a sapphire machining process integrated with intermediate-scale quantum processors. The process allows through-substrate electrical connections, necessary for low-frequency mode-mitigation, as well as signal-routing, which are vital as quantum computers scale in qubit number, and thus dimension. High-coherence qubits are required to build fault-tolerant quantum computers and so material choices are an important consideration when developing a qubit technology platform. Sapphire, as a low-loss dielectric substrate, has shown to support high-coherence qubits. In addition, recent advances in material choices such as tantalum and titanium-nitride, both deposited on a sapphire substrate, have demonstrated qubit lifetimes exceeding 0.3 ms. However, the lack of any process equivalent of deep-silicon etching to create through-substrate-vias in sapphire, or to inductively shunt large dies, has limited sapphire to small-scale processors, or necessitates the use of chiplet architecture. Here, we present a sapphire machining process that is compatible with  high-coherence qubits. This technique immediately provides a means to scale QPUs with integrated mode-mitigation, and provides a route toward the development of through-sapphire-vias, both of which allow the advantages of sapphire to be leveraged as well as facilitating the use of sapphire-compatible materials for large-scale QPUs.
\end{abstract}

\maketitle

%\tableofcontents

\section{\label{sec:intro}Introduction}

To progress toward fault-tolerant quantum computing large numbers of qubits are required. Superconducting qubits present a promising platform with which to scale quantum processors \cite{Arute2019, Bravyi2022, Acharya2023}. As processor qubit number increases, typically the dimensions of the processors also increase. With this scaling, the size of the quantum processing unit (QPU) can become physically large enough that the dimensions of the cavity or enclosure which houses the processor can support modes that are commensurate with the qubit frequencies. Multiple approaches to avoid these spurious modes have been implemented: on typical planar devices separated ground planes can be inductively shunted with airbridges \cite{Chen2014}, or connected with through-substrate-vias (TSVs) \cite{Yost2020, Vahidpour2017, AlfaroBarrantes2020}. Low-frequency cavity modes can be avoided by ensuring that the cavities are small enough such that these modes cannot be supported. To this   end, quantum processors have been divided into chiplets with each chiplet housed in smaller ($\approx1\,\mathrm{cm}$) cavities. The processors can also be housed in inductively shunted cavities \cite{Murray2016, Spring2020}. Spring \textit{et al.} demonstrated an architecture that provides this inductive shunting by means of a conducting pillar passing through an aperture in the substrate and connecting the top and bottom walls of the enclosure \cite{Spring2022}.  TSVs and inductively shunted cavities both require electrical connections to pass through the substrate of the QPU, a critical capability to scale the size of QPUs. 

Despite the technological importance of superconducting qubits, they rely on a small number of critical materials. With few exceptions, high performance qubits have superconducting pads made from Al \cite{Arute2019, Biznarova2023,Osman2023}, Nb \cite{Bal2024}, Ta \cite{Place2021, Wang2022}, TiN \cite{Deng2023} or nitrides of the previously listed metals \cite{Kim2021}. The Josephson junctions are typically made from Al/AlOx/Al tunnel barriers and they are manufactured on high resistivity silicon or sapphire substrates. Although both substrates are currently compatible  with state of the art performance, sapphire is incompatible with large scale integration as there are no processes which allow the through-substrate electrical connections required for mode mitigation. Historically, this has left just one viable substrate material, silicon, which has been used to integrate many complex 3D layered devices \cite{Yost2020, Vahidpour2017, Hazard2023}.

High-coherence qubits have been manufactured on silicon substrates; records include $T_1 \sim 300\,\mu$s using capped Nb films \cite{Bal2024}, $T_1 \sim 200\,\mu$s using uncapped Nb films \cite{Kono2024} and $T_1 \sim 270\,\mu$s using Al films \cite{Biznarova2023}. Whereas on sapphire substrates records include $T_1 = 300 - 400\,\mu$s using TiN films \cite{Deng2023} and $T_1 = 300 - 400\,\mu$s using Ta films \cite{Place2021, Wang2022}. In the case of tantalum, high coherence qubits have only been shown on sapphire. Dielectric loss measurements of sapphire performed at mK show record loss tangents with crystals grown by the heat-exchanger method (HEM) reporting tan$\delta_{\rm bulk} = 1.9\times10^{-8}$ \cite{Read2023}, compared to measurements of dielectric loss on silicon reporting tan$\delta_{\rm bulk} = 2.7\times10^{-6}$ \cite{Checchin2022}. In addition to sapphire currently showing lower loss tangents compared to silicon, acceptor loss mechanisms in silicon may provide a hard-to-engineer loss mechanism \cite{Zhang2024}.

As sapphire offers a low-loss platform for high coherence qubits, a route towards scaling and mitigating the modes that come with increased chip dimension is required. In this work we demonstrate a complete end-to-end manufacturing process of an Oxford Quantum Circuits (OQC) 32-qubit QPU ``Toshiko" integrated with through-sapphire machining to incorporate through-sapphire pillars which inductively shunt the QPU enclosure for mode-mitigation purposes. The demonstration of high-coherence qubits on a sapphire substrate that has undergone a computer numerical control (CNC) machining process effectively unlocks sapphire as a technologically relevant platform to scale superconducting qubits.

\section{\label{sec:level1}Results}

\subsection{\label{sec:level1}Qubit Fabrication and Sapphire Machining Integration}

\begin{figure}
\includegraphics[width=0.5\textwidth]{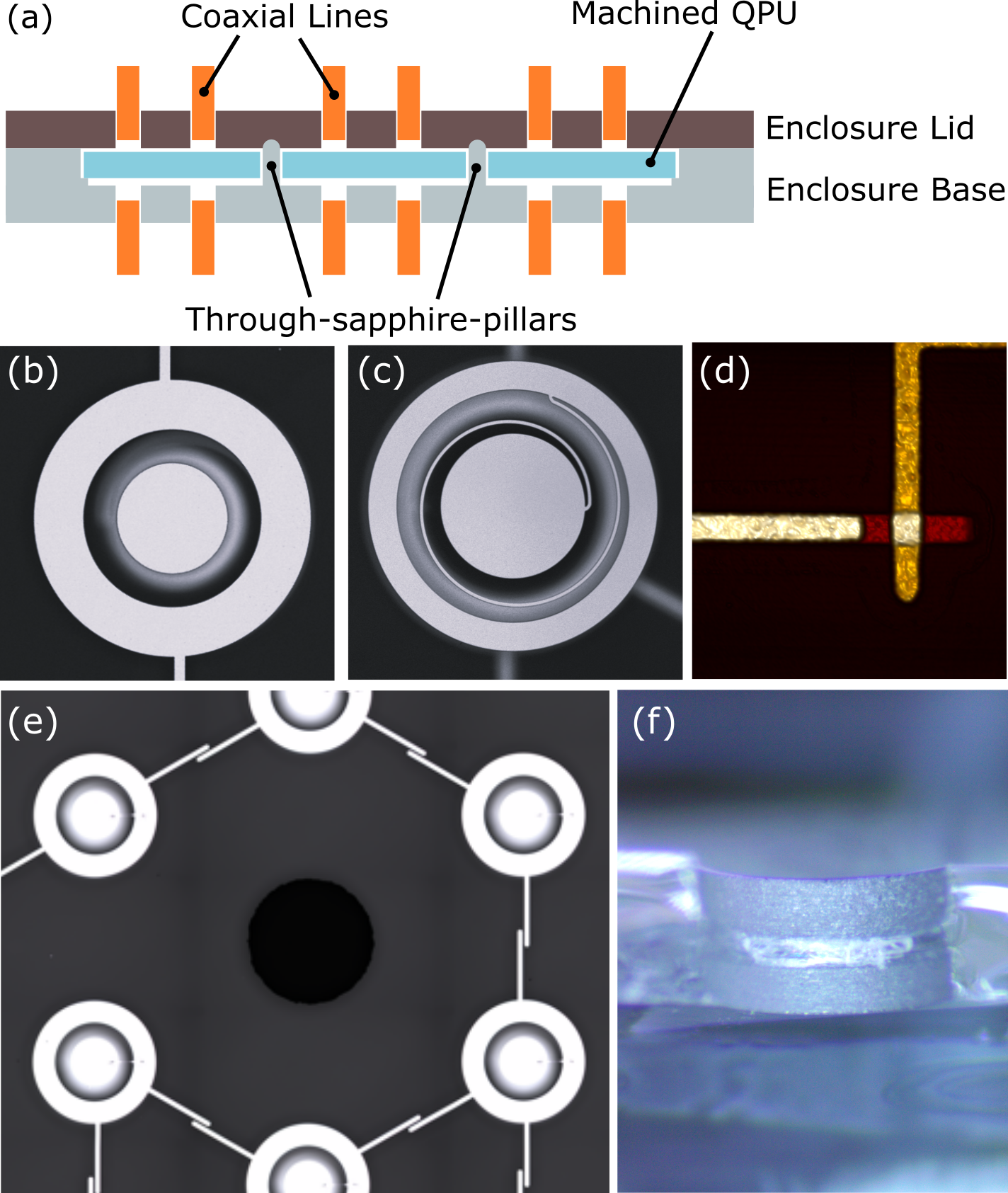}
\caption{\label{fig:fab_overview} (a) Schematic of sapphire QPU with through-substrate machining. Pillars extrude from enclosure base protruding through the sapphire QPU and make electrical contact with the enclosure lid. Constituent components of a substrate-drilled coaxmon-based QPU: (b) The qubit electrodes are fabricated on the top side of the wafer using a combination of e-beam deposition, photolithography and etching processes. (c) The lumped LC spiral resonator is similarly fabricated on the back side of the wafer. (d) An AFM image of the Dolan bridge Josephson junction. The Josephson junction is defined by electron-beam lithography and fabricated using a typical double-angle shadow evaporation technique on the top side of the wafer forming the qubit cell. (e) Shows a micrograph of a drilled aperture in sapphire in close vicinity to the qubit lattice. The drilling is performed after all of the fabrication is complete, but prior to wafer dicing. The qubit and resonator outer diameters are 1\,mm. (f) Shows a profile micrograph of a cleaved sapphire die post-machining (cleaved through the middle of the aperture). The ridge in the middle of the aperture is due to tool wear.}
\end{figure}

The qubits used in this work are coaxmons - an architecture that necessitates fabricating on both sides of a wafer. The coaxmon is an implementation of the transmon whereby the qubit has a coaxial geometry and is fabricated on one side of a substrate \cite{Rahamim2017}. The corresponding readout resonator is a lumped LC spiral resonator and is fabricated on the other side of the substrate aligned to the qubit. Each qubit cell is capacitively coupled to control and readout ports, and may be scaled to large 2D qubit arrays.  See Fig \ref{fig:fab_overview} for examples of each constituent component of the coaxmon.

Integration of TSVs in silicon is usually achieved using high aspect ratio chemical etch process such as deep reactive ion etching (DRIE) \cite{Wu2010}.
As sapphire is inert to most physical and chemical etching, the high aspect ratio etching processes used ubiquitously for silicon TSVs do not exist for sapphire \cite{Grigoras2022}. However, it is possible to create apertures in sapphire using a laser beam in a process known as laser drilling \cite{Schulz2013}. The material removal during laser drilling is an ablative process with a large amount of energy deposited into the substrate which heats the area around the drilling site \cite{Jia2022}. Due to the extent of this heating we have found laser drilling to be difficult to integrate with our manufacturing process. For more details see Section \ref{sec:laser}.
An alternative to etching and laser drilling to create apertures or vias in silicon is CNC machining. Apertures in silicon achieved using CNC machining has been demonstrated on 4-qubit devices \cite{Spring2022}, with qubit coherence times in excess of $100\,\mu\mathrm{s}$.

\begin{figure*}
\includegraphics[width=1\textwidth]{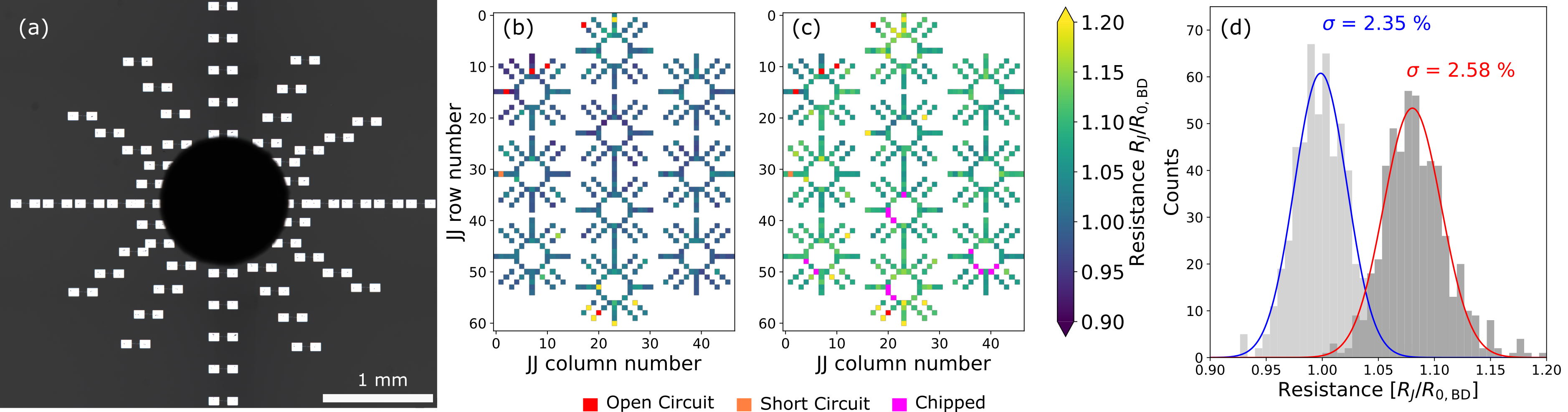}
\caption{\label{fig:drill_overview}The figure demonstrates the robustness of the Josephson junctions to the mechanical drilling process. (a) A micrograph of a radial test pattern of Josephson junctions. A 1\,mm aperture was machined though the sapphire in close vicinity to the junctions. The aperture is surrounded by 56 JJs in a radial pattern. (b) A composite resistance colour-map of 10 drilled apertures before machining. The JJ resistances are measured prior to drilling and any failures such as a short-circuit or an open-circuit are recorded (shown as orange and red on the colourmap respectively). (c) JJ resistance measurements of the same die after aperture machining. Any additional JJ failures are recorded. In addition we record JJs that were unmeasureable due to chipping of the sapphire. (d) Histograms of the JJ resistances prior to, and after, the sapphire machining process. Prior to machining the resistance spread is $\sigma = 2.35\,\%$. After the machining the resistance spread is slightly increased to  $\sigma = 2.58\,\%$. Note that there is also a shift of the die median after machining. This shift in resistance may be due to the time between subsequent measurements (known as aging), as a result of the machining itself, or a combination of both effects.}
\end{figure*}

Despite sapphire’s high hardness and relatively low fracture toughness which makes it challenging to process via physical machining, we successfully demonstrate the use of CNC machining as an alternative to plasma etching processes and laser drilling to create through-sapphire apertures. The apertures were CNC machined in the sapphire using a Loxham Precision $\mu6$ micro-milling system. A diamond micro-grain tool measuring $600\,\mu\mathrm{m}$ in diameter, was used to form 1\,mm diameter apertures following a helical toolpath. During drilling, the substrate and grinding pin are actively cooled using DI water; this keeps the grinding pin cool, but also mitigates the heating of the Josephson junctions and protective coatings. The substrate is coated in a series of charge mitigation and photoresist layers to protect the nanofabricated circuits from electrostatic damage as well as machining debris.
Note that heat affecting the JJs can be avoided by machining the apertures in advance of fabricating the JJs. However, the resulting resist inhomogeneity  due to apertures being present during resist-spinning leads to devices with increased wafer-scale resistance spread. This is in-line with the results of Muthusubramanian \textit{et al}.~\cite{Muthusubramanian_2024} where they performed a study of both Dolan Bridge and Manhattan style Josephson junctions with and without integrated TSVs. They showed that in both styles of JJs the wafer-scale resistance spread is increased on a TSV-integrated substrate and is attributed to an increase in resist-height variation on the TSV-integrated substrates when compared to planar substrates.

\subsection{\label{sec:level1}Josephson Junction Sensitivity to Machining}

To determine the robustness of the Josephson junctions to the drilling process, a test wafer was fabricated with JJs placed in a radial pattern surrounding where the substrate machining would be performed. The pattern was made to determine if there was a higher incidence of failures close to the drilling site, as well as to determine if the machining processes causes any residual heating that can affect the junctions due to heat-induced accelerated aging \cite{Kreikebaum2020}. Figure \ref{fig:drill_overview} (a) shows the radial JJ test dies with an approximately $1\,\mathrm{mm}$ diameter aperture machined in the centre. Figure \ref{fig:drill_overview} (b) and (c) show the resistance colour-maps before and after the substrate machining process, respectively. Failure mechanisms such as open-circuits and short-circuits are recorded. Damage from sapphire chipping is also recorded in the post-machined map (c). The histograms in Figure \ref{fig:drill_overview} (d) confirm that the increase in JJ resistance spread after the drilling process is small ($\sigma = 2.35\,\%$ before drilling and $\sigma = 2.58\,\%$ after drilling). We see no clear spatial dependence of the JJs resistance change in relation to the aperture machining site. The as fabrication JJ yield (JJs not short-circuit or open-circuit) of the devices shown in Fig \ref{fig:drill_overview} (b) was 98.93\,\%. After the drilling 14 JJs could not be measured due to chipping of the pads very close to the drill location. No additional short-circuit or open-circuits were caused by the drilling in this die. (We note that in a co-drilled die an additional 4 short-circuits were measured post-drilling, representing a 99.28\,\% drilling success rate).

\begin{figure*}
\includegraphics[width=\textwidth]{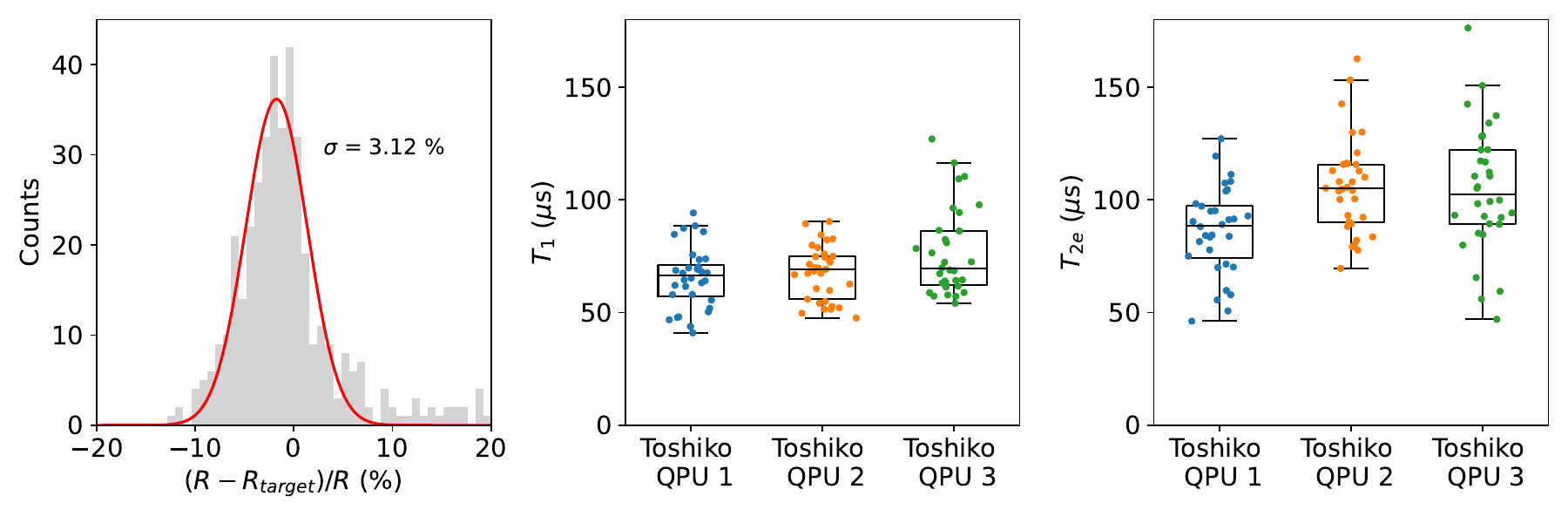}
\caption{\label{fig:coherence_overview}(a) Histogram of Josephson junction resistances measured on a $3\,"$ wafer of Toshiko 32Q QPUs. The histogram is a composite of three sets of JJs normalised to their median, or target value $R_\mathrm{target}$, showing spread in resistance of $3.1\,\%$. The resistances are measured at the end of the full fabrication process inclusive of both sapphire aperture machining, and dicing; thus the spread represents the full end-to-end QPU process. (b) Boxplots of $3\times$ 32Q QPU $T_{1}$ relaxation times. Each circle is a median value for 50+ measurements.  (c) Boxplots of $3\times$ 32Q QPU $T_{2e}$ coherence times. Each circle is a median value for 50+ measurements. characteristic decay times. The full-QPU median values are shown in Table \ref{tab:table1}.}
\end{figure*}

\subsection{\label{sec:level1}Full Wafer QPU Production and  Coherence Measurements}

For our 32Q QPU manufacturing, we use $3\,"$ double-sided-polished sapphire as the substrate. The aperture machining process is performed after fabrication. The wafer is then diced into individual QPUs prior to final cleaning. The electrical resistance of the Josephson junctions are measured using a probe station to ascertain if any of the JJs have become damaged (for instance a short-circuit or open-circuit). Figure \ref{fig:coherence_overview}(a) shows a a histogram of resistance values measured on a representative $3\,"$ wafer of Toshiko 32Q QPUs with a spread of resistance values $\sigma=3.12\,\%$. Note that the resistances presented are measured at the end of the full manufacturing process, inclusive of fabrication, sapphire aperture machining, and dicing. This demonstrates that our process is able to produce wafer-scale Josephson junction spread commensurate with state-of-the-art values whilst incorporating additional machining of the QPU. Recent improvements in our Josephson junction fabrication have reduced this spread to 2.44\% at 3" wafer-scale (see Section \ref{sec:spread-244}) without any JJ post-processing such as laser-tuning \cite{Hertzberg2020}, electron-beam-tuning \cite{Balaji2024}, or voltage-tuning \cite{Pappas2024}.

To demonstrate that the sapphire machining process is compatible with the manufacture of high coherence qubits and QPUs, measurements of $T_{1}$ and $T_{2e}$ coherence statistics for three OQC Toshiko 32Q QPUs are presented. The Toshiko 32Q QPU has eight 1\,mm apertures machined through the sapphire substrate allowing conducting pillars to pass through the substrate to connect the top and bottom of the QPU cavity. This forms the inductive shunting required for cavity mode mitigation as discussed in Sec \ref{sec:intro} \cite{Spring2020, Spring2022}.

The $T_{1}$ relaxation time and $T_{2e}$ coherence times statistics are presented for the Toshiko 32Q QPUs as shown in Figure \ref{fig:coherence_overview}(b) and (c) respectively. As per coherence-reporting best-practice, median values of a statistically significant number of coherence measurements are reported \cite{Burnett2019}. Each data point in the plot represents a single qubit and the value plotted is a median of 50+ decay time measurements. The full-QPU median values are presented in Table \ref{tab:table1}. 
\\
\begin{table}[]
\caption{\label{tab:table1}%
Table presents the $T_{1}$ and $T_{2e}$ coherence data from 3 Toshiko QPUs. Full-QPU median values are shown for each characteristic decay time.}
\begin{ruledtabular}
\begin{tabular}{lcc}
\textrm{QPU}&
\textrm{$T_{1} (\mu\mathrm{s})$}&
\textrm{$T_{2e} (\mu\mathrm{s})$}\\
\colrule
Toshiko-1 & $66.4 \pm 13.5$ & $88.6 \pm 19.5$ \\
Toshiko-2 & $69.2 \pm 12.0$ & $105.2 \pm 21.3$ \\
Toshiko-3 & $69.4 \pm 19.5$ & $102.6 \pm 28.6$\\
\end{tabular}
\end{ruledtabular}
\end{table}
 The single-qubit-median best values are $T_{1}=127\,\mu\mathrm{s}$ and $T_{2e}=176\,\mu\mathrm{s}$. Also notable are the single-qubit-median worst values where $T_{1}=40\,\mu\mathrm{s}$ and $T_{2e}=42\,\mu\mathrm{s}$ representing 60\%, and 47\%, of the ensemble median values of a total of 96 qubits. The absence of qubits with anomalously low coherence is highly desirable for QPUs, as it means that algorithmic chains are not broken by error hot-spots. The coherence times reported here are in-line with those on silicon substrates in pillar-integrated packages at smaller scale as demonstrated in Spring \textit{et al.} \cite{Spring2022}, as well with uncoupled qubit coherence values from OQC 8Q devices as shown in \cite{Balaji2024}. The coherence times are an improvement over OQC's previous generation Lucy 8Q QPU \cite{OQC_Lucy} despite scaling in qubit count, die size, and integrating the additional sapphire machining process steps. As the coherence numbers reported in Fig.~\ref{fig:coherence_overview} are from a Toshiko-generation quantum computer, coupling rates are targeted for aggregate performance i.e. gate and readout fidelities. It is likely that by reducing coupling rates these coherence times could be further improved, showing the ultimate material-platform limit of our end-to-end process and the numbers presented are a lower bound on this.

\section{\label{sec:level1}Discussion}

We have demonstrated a technique to create through-sapphire apertures and have integrated this with a 32-qubit QPU. We demonstrate high yield of the Josephson junctions following the full fabrication flow and the post-fabrication drilling and dicing. We have shown by measurements of $T_{1}$ and $T_{2e}$ coherence statistics for our Toshiko 32Q QPU processor that the mechanical drilling technique is compatible with a high-coherence qubit platform. This work creates an opportunity for the quantum computing field to further explore promising low-loss dielectric substrates such as sapphire. As such, it encourages further development of material platforms such as tantalum (deposited on sapphire) as they now can also be scaled further. Although we demonstrate the through-sapphire apertures for the purpose of inductive shunt package integration, the results are promising for future development of through-sapphire-vias, specifically for the purpose of signal routing and delivery. We demonstrate proof-of-principle vias using electron-beam evaporation in \ref{sec:Supp}. This work also unlocks the potential for sapphire as a low-loss dielectric for QPU-adjacent quantum interposers.

%\clearpage
%\newpage
\section{\label{sec:level1}Methods}
%\subsection{\label{sec:level1}Coaxmon Qubit Fabrication}

\subsection{\label{sec:fab}Fabrication}

The qubits are fabricated on 3" double-sided-polished sapphire. Aluminium is deposited on one side of the wafer and then qubit electrodes are formed using photolithography and etching. The process is repeated on the back side of the wafer to form the coaxial resonators. Finally the Josephson junctions are fabricated using a typical Dolan Bridge double-angle shadow evaporation deposition \cite{Dolan1977}. The QPUs were then machined using the Loxham CNC before being diced into individual QPU dies.

\begin{acknowledgments}
We extend our thanks to the entire OQC Team for their contributions to the quantum computing stack, which was instrumental in this work. Specifically we acknowledge the OQC Toshiko Team for the coherence measurements used as part of this work. Thanks to Jonathan Burnett and Brian Vlastakis for their review of this manuscript.  We thank Rais Shaikhaidarov and Phil Meeson for their support at the Royal Holloway University of London SuperFab Facility. We thank Nadeem Rizvi and Laser Micromachining Ltd for useful discussions around laser drilling as well as carrying out the laser drilling. PGC and JCG would like to acknowledge the following funding from the Engineering and Physical Sciences Research Council (EP/M013243/1, EP/T001062/1, and EP/W024772/1).
\end{acknowledgments}

\subsection{\label{sec:author-cont}Author Contributions}

KS, NA, YB, RFA, KGC, and RDP fabricated the devices and QPUs, carried out the JJ resistance measurements, and performed the microscopy. PCG and JCG developed the sapphire machining process. KS, NA, YB, RFA, KGC, RDP, OWK, and CDS developed the process flow to integrate the aperture machining with QPUs and JJs. PCG carried out the aperture machining. YB and CDS oversaw machining of the QPUs and JJs. KS, OWK, and CDS performed the analysis on the data. KS developed and fabricated the aperture metalisation via proof-of-prinicple. CDS wrote the manuscript with significant input from KS and OWK. CDS led the project.

\clearpage
\newpage
\section{\label{sec:Supp}Supplemental Material}

\subsection{\label{sec:spread-244}Full-wafer JJ Spread}

Through recent fabrication improvements our 3" wafer-scale spread as fabricated, inclusive of sapphire machining and dicing, is as low as 2.44\% - in-line with the best reported as-fabricated values in the literature (for instance see \cite{Osman2023}). Note that qubits from this wafer have not been cryogenically tested so no coherence values are reported. The total yield of this wafer was approximately 80\% due to ESD failures.

\begin{figure}
\includegraphics[width=0.5\textwidth]{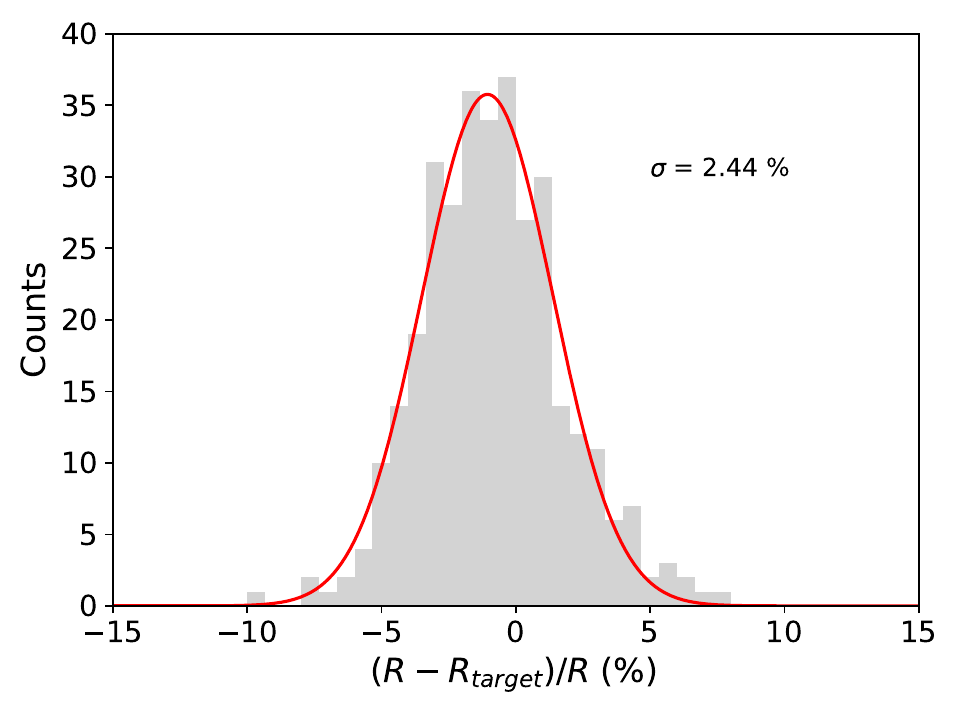}
\caption{\label{fig:spread244}Histogram of Josephson junction resistances measured on a $3\,"$ wafer of Toshiko 32Q QPUs. The histogram is a composite of three sets of JJs normalised to their median, or target value $R_\mathrm{target}$. The composite spread in resistance is $2.44\,\%$}
\end{figure}

\subsection{\label{sec:vias}Demonstration of through-sapphire-vias}

We also demonstrate the metalisation of the through-sapphire-apertures to produce proof-of-principle through-sapphire-vias. The apertures are metalised from both sides using an electron-beam deposition tool to deposit approximately $300\,\mathrm{nm}$ of aluminium through the apertures. An argon milling step is included between the two depositions to ensure ohmic contact, as vacuum is broken in this process. The substrate is rotated throughout the deposition to ensure full aperture wall coverage. Two-probe room temperature resistance measurements between the top and bottom ground planes show a resistance of $7\,\Omega$ demonstrating conduction through the via. Alternative methods of metalising may include means such as sputtering, or by cold-metal injection \cite{Ahmad2014}.

\begin{figure}
\includegraphics[width=0.5\textwidth]{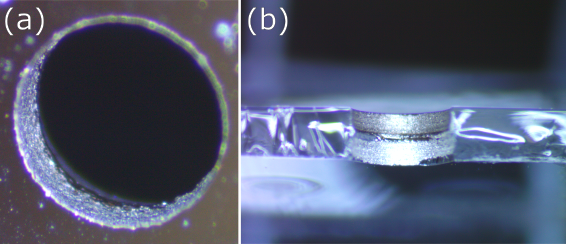}
\caption{\label{fig:vias}(a) Micrograph of 1\,mm diameter drilled aperture with aluminium deposited via electron-beam deposition onto the aperture sidewall. (b) Micrograph of a cleaved sapphire substrate showing the sidewalls of the  1\,mm diameter drilled apertures post aluminium metal deposition. A two-probe resistance measurement showed a $7\,\Omega$ resistance through the via.}
\end{figure}

\subsection{\label{sec:laser}Laser Drilling Integration}

\begin{figure}
\includegraphics[width=0.5\textwidth]{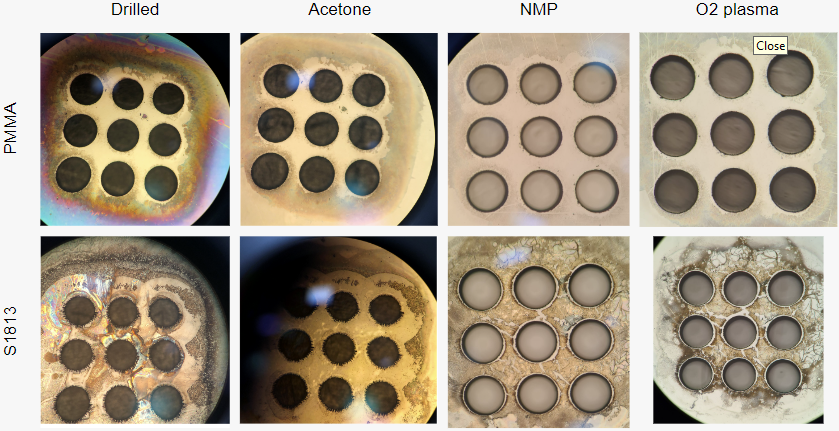}
\caption{\label{fig:laser}(a) Optical micrograph of electron-beam resist (top row) and photoresist (bottom row)  after laser drilling a 3x3 matrix of apertures in the sapphire substrate. The figures show the resist residue remaining after various cleans - acetone (column 2), NMP (column 3) and O2 plasma (column 4). In each case residue remains and cannot be completely removed.}
\end{figure}

It is possible to drill apertures in sapphire using a laser drilling process. As this technique is ablative, the substrate will get hot during the drilling.
This thermal effect can be a limitation for substrates with devices that are affected by heat (for quantum circuits this is important as heat can change the parameters of the Josephson junction). In addition, if any resist is used for protection or further processing then this can lead to thermal cross-linking of the resist causing difficulty in removal. 
In order to integrate this to a process flow in which the JJs are already manufactured protective resist is required. Figure \ref{fig:laser} shows that the laser drilling results in thermal cross-linking of both PMMA and S1813 resists. Attempts at removal using Acetone, NMP, and finally O2 plasma etching were unsuccessful. As resist residue is detrimental to qubit coherence this is not a favourable process for the integration of through-sapphire apertures in quantum circuits \cite{Lisenfeld2019}. In addition, the large-distance heating will result in JJ resistance changes and thus further difficulty with precise frequency allocations in QPUs. 

\clearpage
\newpage

\bibliography{Sapphire-drilling-v1}% Produces the bibliography via BibTeX.

\providecommand{\noopsort}[1]{}\providecommand{\singleletter}[1]{#1}%
\begin{thebibliography}{37}%
\makeatletter
\providecommand \@ifxundefined [1]{%
 \@ifx{#1\undefined}
}%
\providecommand \@ifnum [1]{%
 \ifnum #1\expandafter \@firstoftwo
 \else \expandafter \@secondoftwo
 \fi
}%
\providecommand \@ifx [1]{%
 \ifx #1\expandafter \@firstoftwo
 \else \expandafter \@secondoftwo
 \fi
}%
\providecommand \natexlab [1]{#1}%
\providecommand \enquote  [1]{``#1''}%
\providecommand \bibnamefont  [1]{#1}%
\providecommand \bibfnamefont [1]{#1}%
\providecommand \citenamefont [1]{#1}%
\providecommand \href@noop [0]{\@secondoftwo}%
\providecommand \href [0]{\begingroup \@sanitize@url \@href}%
\providecommand \@href[1]{\@@startlink{#1}\@@href}%
\providecommand \@@href[1]{\endgroup#1\@@endlink}%
\providecommand \@sanitize@url [0]{\catcode `\\12\catcode `\$12\catcode `\&12\catcode `\#12\catcode `\^12\catcode `\_12\catcode `\%12\relax}%
\providecommand \@@startlink[1]{}%
\providecommand \@@endlink[0]{}%
\providecommand \url  [0]{\begingroup\@sanitize@url \@url }%
\providecommand \@url [1]{\endgroup\@href {#1}{\urlprefix }}%
\providecommand \urlprefix  [0]{URL }%
\providecommand \Eprint [0]{\href }%
\providecommand \doibase [0]{https://doi.org/}%
\providecommand \selectlanguage [0]{\@gobble}%
\providecommand \bibinfo  [0]{\@secondoftwo}%
\providecommand \bibfield  [0]{\@secondoftwo}%
\providecommand \translation [1]{[#1]}%
\providecommand \BibitemOpen [0]{}%
\providecommand \bibitemStop [0]{}%
\providecommand \bibitemNoStop [0]{.\EOS\space}%
\providecommand \EOS [0]{\spacefactor3000\relax}%
\providecommand \BibitemShut  [1]{\csname bibitem#1\endcsname}%
\let\auto@bib@innerbib\@empty
%</preamble>
\bibitem [{\citenamefont {Arute}\ \emph {et~al.}(2019)\citenamefont {Arute}, \citenamefont {Arya}, \citenamefont {Babbush}, \citenamefont {Bacon}, \citenamefont {Bardin}, \citenamefont {Barends}, \citenamefont {Biswas}, \citenamefont {Boixo}, \citenamefont {Brandao}, \citenamefont {Buell}, \citenamefont {Burkett}, \citenamefont {Chen}, \citenamefont {Chen}, \citenamefont {Chiaro}, \citenamefont {Collins}, \citenamefont {Courtney}, \citenamefont {Dunsworth}, \citenamefont {Farhi}, \citenamefont {Foxen}, \citenamefont {Fowler}, \citenamefont {Gidney}, \citenamefont {Giustina}, \citenamefont {Graff}, \citenamefont {Guerin}, \citenamefont {Habegger}, \citenamefont {Harrigan}, \citenamefont {Hartmann}, \citenamefont {Ho}, \citenamefont {Hoffmann}, \citenamefont {Huang}, \citenamefont {Humble}, \citenamefont {Isakov}, \citenamefont {Jeffrey}, \citenamefont {Jiang}, \citenamefont {Kafri}, \citenamefont {Kechedzhi}, \citenamefont {Kelly}, \citenamefont {Klimov}, \citenamefont {Knysh}, \citenamefont {Korotkov},
  \citenamefont {Kostritsa}, \citenamefont {Landhuis}, \citenamefont {Lindmark}, \citenamefont {Lucero}, \citenamefont {Lyakh}, \citenamefont {Mandr{\`a}}, \citenamefont {McClean}, \citenamefont {McEwen}, \citenamefont {Megrant}, \citenamefont {Mi}, \citenamefont {Michielsen}, \citenamefont {Mohseni}, \citenamefont {Mutus}, \citenamefont {Naaman}, \citenamefont {Neeley}, \citenamefont {Neill}, \citenamefont {Niu}, \citenamefont {Ostby}, \citenamefont {Petukhov}, \citenamefont {Platt}, \citenamefont {Quintana}, \citenamefont {Rieffel}, \citenamefont {Roushan}, \citenamefont {Rubin}, \citenamefont {Sank}, \citenamefont {Satzinger}, \citenamefont {Smelyanskiy}, \citenamefont {Sung}, \citenamefont {Trevithick}, \citenamefont {Vainsencher}, \citenamefont {Villalonga}, \citenamefont {White}, \citenamefont {Yao}, \citenamefont {Yeh}, \citenamefont {Zalcman}, \citenamefont {Neven},\ and\ \citenamefont {Martinis}}]{Arute2019}%
  \BibitemOpen
  \bibfield  {author} {\bibinfo {author} {\bibfnamefont {F.}~\bibnamefont {Arute}}, \bibinfo {author} {\bibfnamefont {K.}~\bibnamefont {Arya}}, \bibinfo {author} {\bibfnamefont {R.}~\bibnamefont {Babbush}}, \bibinfo {author} {\bibfnamefont {D.}~\bibnamefont {Bacon}}, \bibinfo {author} {\bibfnamefont {J.~C.}\ \bibnamefont {Bardin}}, \bibinfo {author} {\bibfnamefont {R.}~\bibnamefont {Barends}}, \bibinfo {author} {\bibfnamefont {R.}~\bibnamefont {Biswas}}, \bibinfo {author} {\bibfnamefont {S.}~\bibnamefont {Boixo}}, \bibinfo {author} {\bibfnamefont {F.~G. S.~L.}\ \bibnamefont {Brandao}}, \bibinfo {author} {\bibfnamefont {D.~A.}\ \bibnamefont {Buell}}, \bibinfo {author} {\bibfnamefont {B.}~\bibnamefont {Burkett}}, \bibinfo {author} {\bibfnamefont {Y.}~\bibnamefont {Chen}}, \bibinfo {author} {\bibfnamefont {Z.}~\bibnamefont {Chen}}, \bibinfo {author} {\bibfnamefont {B.}~\bibnamefont {Chiaro}}, \bibinfo {author} {\bibfnamefont {R.}~\bibnamefont {Collins}}, \bibinfo {author} {\bibfnamefont {W.}~\bibnamefont
  {Courtney}}, \bibinfo {author} {\bibfnamefont {A.}~\bibnamefont {Dunsworth}}, \bibinfo {author} {\bibfnamefont {E.}~\bibnamefont {Farhi}}, \bibinfo {author} {\bibfnamefont {B.}~\bibnamefont {Foxen}}, \bibinfo {author} {\bibfnamefont {A.}~\bibnamefont {Fowler}}, \bibinfo {author} {\bibfnamefont {C.}~\bibnamefont {Gidney}}, \bibinfo {author} {\bibfnamefont {M.}~\bibnamefont {Giustina}}, \bibinfo {author} {\bibfnamefont {R.}~\bibnamefont {Graff}}, \bibinfo {author} {\bibfnamefont {K.}~\bibnamefont {Guerin}}, \bibinfo {author} {\bibfnamefont {S.}~\bibnamefont {Habegger}}, \bibinfo {author} {\bibfnamefont {M.~P.}\ \bibnamefont {Harrigan}}, \bibinfo {author} {\bibfnamefont {M.~J.}\ \bibnamefont {Hartmann}}, \bibinfo {author} {\bibfnamefont {A.}~\bibnamefont {Ho}}, \bibinfo {author} {\bibfnamefont {M.}~\bibnamefont {Hoffmann}}, \bibinfo {author} {\bibfnamefont {T.}~\bibnamefont {Huang}}, \bibinfo {author} {\bibfnamefont {T.~S.}\ \bibnamefont {Humble}}, \bibinfo {author} {\bibfnamefont {S.~V.}\ \bibnamefont
  {Isakov}}, \bibinfo {author} {\bibfnamefont {E.}~\bibnamefont {Jeffrey}}, \bibinfo {author} {\bibfnamefont {Z.}~\bibnamefont {Jiang}}, \bibinfo {author} {\bibfnamefont {D.}~\bibnamefont {Kafri}}, \bibinfo {author} {\bibfnamefont {K.}~\bibnamefont {Kechedzhi}}, \bibinfo {author} {\bibfnamefont {J.}~\bibnamefont {Kelly}}, \bibinfo {author} {\bibfnamefont {P.~V.}\ \bibnamefont {Klimov}}, \bibinfo {author} {\bibfnamefont {S.}~\bibnamefont {Knysh}}, \bibinfo {author} {\bibfnamefont {A.}~\bibnamefont {Korotkov}}, \bibinfo {author} {\bibfnamefont {F.}~\bibnamefont {Kostritsa}}, \bibinfo {author} {\bibfnamefont {D.}~\bibnamefont {Landhuis}}, \bibinfo {author} {\bibfnamefont {M.}~\bibnamefont {Lindmark}}, \bibinfo {author} {\bibfnamefont {E.}~\bibnamefont {Lucero}}, \bibinfo {author} {\bibfnamefont {D.}~\bibnamefont {Lyakh}}, \bibinfo {author} {\bibfnamefont {S.}~\bibnamefont {Mandr{\`a}}}, \bibinfo {author} {\bibfnamefont {J.~R.}\ \bibnamefont {McClean}}, \bibinfo {author} {\bibfnamefont {M.}~\bibnamefont
  {McEwen}}, \bibinfo {author} {\bibfnamefont {A.}~\bibnamefont {Megrant}}, \bibinfo {author} {\bibfnamefont {X.}~\bibnamefont {Mi}}, \bibinfo {author} {\bibfnamefont {K.}~\bibnamefont {Michielsen}}, \bibinfo {author} {\bibfnamefont {M.}~\bibnamefont {Mohseni}}, \bibinfo {author} {\bibfnamefont {J.}~\bibnamefont {Mutus}}, \bibinfo {author} {\bibfnamefont {O.}~\bibnamefont {Naaman}}, \bibinfo {author} {\bibfnamefont {M.}~\bibnamefont {Neeley}}, \bibinfo {author} {\bibfnamefont {C.}~\bibnamefont {Neill}}, \bibinfo {author} {\bibfnamefont {M.~Y.}\ \bibnamefont {Niu}}, \bibinfo {author} {\bibfnamefont {E.}~\bibnamefont {Ostby}}, \bibinfo {author} {\bibfnamefont {A.}~\bibnamefont {Petukhov}}, \bibinfo {author} {\bibfnamefont {J.~C.}\ \bibnamefont {Platt}}, \bibinfo {author} {\bibfnamefont {C.}~\bibnamefont {Quintana}}, \bibinfo {author} {\bibfnamefont {E.~G.}\ \bibnamefont {Rieffel}}, \bibinfo {author} {\bibfnamefont {P.}~\bibnamefont {Roushan}}, \bibinfo {author} {\bibfnamefont {N.~C.}\ \bibnamefont {Rubin}},
  \bibinfo {author} {\bibfnamefont {D.}~\bibnamefont {Sank}}, \bibinfo {author} {\bibfnamefont {K.~J.}\ \bibnamefont {Satzinger}}, \bibinfo {author} {\bibfnamefont {V.}~\bibnamefont {Smelyanskiy}}, \bibinfo {author} {\bibfnamefont {K.~J.}\ \bibnamefont {Sung}}, \bibinfo {author} {\bibfnamefont {M.~D.}\ \bibnamefont {Trevithick}}, \bibinfo {author} {\bibfnamefont {A.}~\bibnamefont {Vainsencher}}, \bibinfo {author} {\bibfnamefont {B.}~\bibnamefont {Villalonga}}, \bibinfo {author} {\bibfnamefont {T.}~\bibnamefont {White}}, \bibinfo {author} {\bibfnamefont {Z.~J.}\ \bibnamefont {Yao}}, \bibinfo {author} {\bibfnamefont {P.}~\bibnamefont {Yeh}}, \bibinfo {author} {\bibfnamefont {A.}~\bibnamefont {Zalcman}}, \bibinfo {author} {\bibfnamefont {H.}~\bibnamefont {Neven}},\ and\ \bibinfo {author} {\bibfnamefont {J.~M.}\ \bibnamefont {Martinis}},\ }\bibfield  {title} {\bibinfo {title} {Quantum supremacy using a programmable superconducting processor},\ }\href {https://doi.org/10.1038/s41586-019-1666-5} {\bibfield
  {journal} {\bibinfo  {journal} {Nature}\ }\textbf {\bibinfo {volume} {574}},\ \bibinfo {pages} {505} (\bibinfo {year} {2019})}\BibitemShut {NoStop}%
\bibitem [{\citenamefont {Bravyi}\ \emph {et~al.}(2022)\citenamefont {Bravyi}, \citenamefont {Dial}, \citenamefont {Gambetta}, \citenamefont {Gil},\ and\ \citenamefont {Nazario}}]{Bravyi2022}%
  \BibitemOpen
  \bibfield  {author} {\bibinfo {author} {\bibfnamefont {S.}~\bibnamefont {Bravyi}}, \bibinfo {author} {\bibfnamefont {O.}~\bibnamefont {Dial}}, \bibinfo {author} {\bibfnamefont {J.~M.}\ \bibnamefont {Gambetta}}, \bibinfo {author} {\bibfnamefont {D.}~\bibnamefont {Gil}},\ and\ \bibinfo {author} {\bibfnamefont {Z.}~\bibnamefont {Nazario}},\ }\bibfield  {title} {\bibinfo {title} {The future of quantum computing with superconducting qubits},\ }\href {https://doi.org/10.1063/5.0082975} {\bibfield  {journal} {\bibinfo  {journal} {Journal of Applied Physics}\ }\textbf {\bibinfo {volume} {132}},\ \bibinfo {pages} {160902} (\bibinfo {year} {2022})}\BibitemShut {NoStop}%
\bibitem [{\citenamefont {Acharya}\ \emph {et~al.}(2023)\citenamefont {Acharya}, \citenamefont {Aleiner}, \citenamefont {Allen}, \citenamefont {Andersen}, \citenamefont {Ansmann}, \citenamefont {Arute}, \citenamefont {Arya}, \citenamefont {Asfaw}, \citenamefont {Atalaya}, \citenamefont {Babbush}, \citenamefont {Bacon}, \citenamefont {Bardin}, \citenamefont {Basso}, \citenamefont {Bengtsson}, \citenamefont {Boixo}, \citenamefont {Bortoli}, \citenamefont {Bourassa}, \citenamefont {Bovaird}, \citenamefont {Brill}, \citenamefont {Broughton}, \citenamefont {Buckley}, \citenamefont {Buell}, \citenamefont {Burger}, \citenamefont {Burkett}, \citenamefont {Bushnell}, \citenamefont {Chen}, \citenamefont {Chen}, \citenamefont {Chiaro}, \citenamefont {Cogan}, \citenamefont {Collins}, \citenamefont {Conner}, \citenamefont {Courtney}, \citenamefont {Crook}, \citenamefont {Curtin}, \citenamefont {Debroy}, \citenamefont {Del Toro~Barba}, \citenamefont {Demura}, \citenamefont {Dunsworth}, \citenamefont {Eppens}, \citenamefont
  {Erickson}, \citenamefont {Faoro}, \citenamefont {Farhi}, \citenamefont {Fatemi}, \citenamefont {Flores~Burgos}, \citenamefont {Forati}, \citenamefont {Fowler}, \citenamefont {Foxen}, \citenamefont {Giang}, \citenamefont {Gidney}, \citenamefont {Gilboa}, \citenamefont {Giustina}, \citenamefont {Grajales~Dau}, \citenamefont {Gross}, \citenamefont {Habegger}, \citenamefont {Hamilton}, \citenamefont {Harrigan}, \citenamefont {Harrington}, \citenamefont {Higgott}, \citenamefont {Hilton}, \citenamefont {Hoffmann}, \citenamefont {Hong}, \citenamefont {Huang}, \citenamefont {Huff}, \citenamefont {Huggins}, \citenamefont {Ioffe}, \citenamefont {Isakov}, \citenamefont {Iveland}, \citenamefont {Jeffrey}, \citenamefont {Jiang}, \citenamefont {Jones}, \citenamefont {Juhas}, \citenamefont {Kafri}, \citenamefont {Kechedzhi}, \citenamefont {Kelly}, \citenamefont {Khattar}, \citenamefont {Khezri}, \citenamefont {Kieferov{\'a}}, \citenamefont {Kim}, \citenamefont {Kitaev}, \citenamefont {Klimov}, \citenamefont {Klots},
  \citenamefont {Korotkov}, \citenamefont {Kostritsa}, \citenamefont {Kreikebaum}, \citenamefont {Landhuis}, \citenamefont {Laptev}, \citenamefont {Lau}, \citenamefont {Laws}, \citenamefont {Lee}, \citenamefont {Lee}, \citenamefont {Lester}, \citenamefont {Lill}, \citenamefont {Liu}, \citenamefont {Locharla}, \citenamefont {Lucero}, \citenamefont {Malone}, \citenamefont {Marshall}, \citenamefont {Martin}, \citenamefont {McClean}, \citenamefont {McCourt}, \citenamefont {McEwen}, \citenamefont {Megrant}, \citenamefont {Meurer~Costa}, \citenamefont {Mi}, \citenamefont {Miao}, \citenamefont {Mohseni}, \citenamefont {Montazeri}, \citenamefont {Morvan}, \citenamefont {Mount}, \citenamefont {Mruczkiewicz}, \citenamefont {Naaman}, \citenamefont {Neeley}, \citenamefont {Neill}, \citenamefont {Nersisyan}, \citenamefont {Neven}, \citenamefont {Newman}, \citenamefont {Ng}, \citenamefont {Nguyen}, \citenamefont {Nguyen}, \citenamefont {Niu}, \citenamefont {O'Brien}, \citenamefont {Opremcak}, \citenamefont {Platt},
  \citenamefont {Petukhov}, \citenamefont {Potter}, \citenamefont {Pryadko}, \citenamefont {Quintana}, \citenamefont {Roushan}, \citenamefont {Rubin}, \citenamefont {Saei}, \citenamefont {Sank}, \citenamefont {Sankaragomathi}, \citenamefont {Satzinger}, \citenamefont {Schurkus}, \citenamefont {Schuster}, \citenamefont {Shearn}, \citenamefont {Shorter}, \citenamefont {Shvarts}, \citenamefont {Skruzny}, \citenamefont {Smelyanskiy}, \citenamefont {Smith}, \citenamefont {Sterling}, \citenamefont {Strain}, \citenamefont {Szalay}, \citenamefont {Torres}, \citenamefont {Vidal}, \citenamefont {Villalonga}, \citenamefont {Vollgraff~Heidweiller}, \citenamefont {White}, \citenamefont {Xing}, \citenamefont {Yao}, \citenamefont {Yeh}, \citenamefont {Yoo}, \citenamefont {Young}, \citenamefont {Zalcman}, \citenamefont {Zhang}, \citenamefont {Zhu},\ and\ \citenamefont {AI}}]{Acharya2023}%
  \BibitemOpen
  \bibfield  {author} {\bibinfo {author} {\bibfnamefont {R.}~\bibnamefont {Acharya}}, \bibinfo {author} {\bibfnamefont {I.}~\bibnamefont {Aleiner}}, \bibinfo {author} {\bibfnamefont {R.}~\bibnamefont {Allen}}, \bibinfo {author} {\bibfnamefont {T.~I.}\ \bibnamefont {Andersen}}, \bibinfo {author} {\bibfnamefont {M.}~\bibnamefont {Ansmann}}, \bibinfo {author} {\bibfnamefont {F.}~\bibnamefont {Arute}}, \bibinfo {author} {\bibfnamefont {K.}~\bibnamefont {Arya}}, \bibinfo {author} {\bibfnamefont {A.}~\bibnamefont {Asfaw}}, \bibinfo {author} {\bibfnamefont {J.}~\bibnamefont {Atalaya}}, \bibinfo {author} {\bibfnamefont {R.}~\bibnamefont {Babbush}}, \bibinfo {author} {\bibfnamefont {D.}~\bibnamefont {Bacon}}, \bibinfo {author} {\bibfnamefont {J.~C.}\ \bibnamefont {Bardin}}, \bibinfo {author} {\bibfnamefont {J.}~\bibnamefont {Basso}}, \bibinfo {author} {\bibfnamefont {A.}~\bibnamefont {Bengtsson}}, \bibinfo {author} {\bibfnamefont {S.}~\bibnamefont {Boixo}}, \bibinfo {author} {\bibfnamefont {G.}~\bibnamefont
  {Bortoli}}, \bibinfo {author} {\bibfnamefont {A.}~\bibnamefont {Bourassa}}, \bibinfo {author} {\bibfnamefont {J.}~\bibnamefont {Bovaird}}, \bibinfo {author} {\bibfnamefont {L.}~\bibnamefont {Brill}}, \bibinfo {author} {\bibfnamefont {M.}~\bibnamefont {Broughton}}, \bibinfo {author} {\bibfnamefont {B.~B.}\ \bibnamefont {Buckley}}, \bibinfo {author} {\bibfnamefont {D.~A.}\ \bibnamefont {Buell}}, \bibinfo {author} {\bibfnamefont {T.}~\bibnamefont {Burger}}, \bibinfo {author} {\bibfnamefont {B.}~\bibnamefont {Burkett}}, \bibinfo {author} {\bibfnamefont {N.}~\bibnamefont {Bushnell}}, \bibinfo {author} {\bibfnamefont {Y.}~\bibnamefont {Chen}}, \bibinfo {author} {\bibfnamefont {Z.}~\bibnamefont {Chen}}, \bibinfo {author} {\bibfnamefont {B.}~\bibnamefont {Chiaro}}, \bibinfo {author} {\bibfnamefont {J.}~\bibnamefont {Cogan}}, \bibinfo {author} {\bibfnamefont {R.}~\bibnamefont {Collins}}, \bibinfo {author} {\bibfnamefont {P.}~\bibnamefont {Conner}}, \bibinfo {author} {\bibfnamefont {W.}~\bibnamefont {Courtney}},
  \bibinfo {author} {\bibfnamefont {A.~L.}\ \bibnamefont {Crook}}, \bibinfo {author} {\bibfnamefont {B.}~\bibnamefont {Curtin}}, \bibinfo {author} {\bibfnamefont {D.~M.}\ \bibnamefont {Debroy}}, \bibinfo {author} {\bibfnamefont {A.}~\bibnamefont {Del Toro~Barba}}, \bibinfo {author} {\bibfnamefont {S.}~\bibnamefont {Demura}}, \bibinfo {author} {\bibfnamefont {A.}~\bibnamefont {Dunsworth}}, \bibinfo {author} {\bibfnamefont {D.}~\bibnamefont {Eppens}}, \bibinfo {author} {\bibfnamefont {C.}~\bibnamefont {Erickson}}, \bibinfo {author} {\bibfnamefont {L.}~\bibnamefont {Faoro}}, \bibinfo {author} {\bibfnamefont {E.}~\bibnamefont {Farhi}}, \bibinfo {author} {\bibfnamefont {R.}~\bibnamefont {Fatemi}}, \bibinfo {author} {\bibfnamefont {L.}~\bibnamefont {Flores~Burgos}}, \bibinfo {author} {\bibfnamefont {E.}~\bibnamefont {Forati}}, \bibinfo {author} {\bibfnamefont {A.~G.}\ \bibnamefont {Fowler}}, \bibinfo {author} {\bibfnamefont {B.}~\bibnamefont {Foxen}}, \bibinfo {author} {\bibfnamefont {W.}~\bibnamefont {Giang}},
  \bibinfo {author} {\bibfnamefont {C.}~\bibnamefont {Gidney}}, \bibinfo {author} {\bibfnamefont {D.}~\bibnamefont {Gilboa}}, \bibinfo {author} {\bibfnamefont {M.}~\bibnamefont {Giustina}}, \bibinfo {author} {\bibfnamefont {A.}~\bibnamefont {Grajales~Dau}}, \bibinfo {author} {\bibfnamefont {J.~A.}\ \bibnamefont {Gross}}, \bibinfo {author} {\bibfnamefont {S.}~\bibnamefont {Habegger}}, \bibinfo {author} {\bibfnamefont {M.~C.}\ \bibnamefont {Hamilton}}, \bibinfo {author} {\bibfnamefont {M.~P.}\ \bibnamefont {Harrigan}}, \bibinfo {author} {\bibfnamefont {S.~D.}\ \bibnamefont {Harrington}}, \bibinfo {author} {\bibfnamefont {O.}~\bibnamefont {Higgott}}, \bibinfo {author} {\bibfnamefont {J.}~\bibnamefont {Hilton}}, \bibinfo {author} {\bibfnamefont {M.}~\bibnamefont {Hoffmann}}, \bibinfo {author} {\bibfnamefont {S.}~\bibnamefont {Hong}}, \bibinfo {author} {\bibfnamefont {T.}~\bibnamefont {Huang}}, \bibinfo {author} {\bibfnamefont {A.}~\bibnamefont {Huff}}, \bibinfo {author} {\bibfnamefont {W.~J.}\ \bibnamefont
  {Huggins}}, \bibinfo {author} {\bibfnamefont {L.~B.}\ \bibnamefont {Ioffe}}, \bibinfo {author} {\bibfnamefont {S.~V.}\ \bibnamefont {Isakov}}, \bibinfo {author} {\bibfnamefont {J.}~\bibnamefont {Iveland}}, \bibinfo {author} {\bibfnamefont {E.}~\bibnamefont {Jeffrey}}, \bibinfo {author} {\bibfnamefont {Z.}~\bibnamefont {Jiang}}, \bibinfo {author} {\bibfnamefont {C.}~\bibnamefont {Jones}}, \bibinfo {author} {\bibfnamefont {P.}~\bibnamefont {Juhas}}, \bibinfo {author} {\bibfnamefont {D.}~\bibnamefont {Kafri}}, \bibinfo {author} {\bibfnamefont {K.}~\bibnamefont {Kechedzhi}}, \bibinfo {author} {\bibfnamefont {J.}~\bibnamefont {Kelly}}, \bibinfo {author} {\bibfnamefont {T.}~\bibnamefont {Khattar}}, \bibinfo {author} {\bibfnamefont {M.}~\bibnamefont {Khezri}}, \bibinfo {author} {\bibfnamefont {M.}~\bibnamefont {Kieferov{\'a}}}, \bibinfo {author} {\bibfnamefont {S.}~\bibnamefont {Kim}}, \bibinfo {author} {\bibfnamefont {A.}~\bibnamefont {Kitaev}}, \bibinfo {author} {\bibfnamefont {P.~V.}\ \bibnamefont {Klimov}},
  \bibinfo {author} {\bibfnamefont {A.~R.}\ \bibnamefont {Klots}}, \bibinfo {author} {\bibfnamefont {A.~N.}\ \bibnamefont {Korotkov}}, \bibinfo {author} {\bibfnamefont {F.}~\bibnamefont {Kostritsa}}, \bibinfo {author} {\bibfnamefont {J.~M.}\ \bibnamefont {Kreikebaum}}, \bibinfo {author} {\bibfnamefont {D.}~\bibnamefont {Landhuis}}, \bibinfo {author} {\bibfnamefont {P.}~\bibnamefont {Laptev}}, \bibinfo {author} {\bibfnamefont {K.-M.}\ \bibnamefont {Lau}}, \bibinfo {author} {\bibfnamefont {L.}~\bibnamefont {Laws}}, \bibinfo {author} {\bibfnamefont {J.}~\bibnamefont {Lee}}, \bibinfo {author} {\bibfnamefont {K.}~\bibnamefont {Lee}}, \bibinfo {author} {\bibfnamefont {B.~J.}\ \bibnamefont {Lester}}, \bibinfo {author} {\bibfnamefont {A.}~\bibnamefont {Lill}}, \bibinfo {author} {\bibfnamefont {W.}~\bibnamefont {Liu}}, \bibinfo {author} {\bibfnamefont {A.}~\bibnamefont {Locharla}}, \bibinfo {author} {\bibfnamefont {E.}~\bibnamefont {Lucero}}, \bibinfo {author} {\bibfnamefont {F.~D.}\ \bibnamefont {Malone}}, \bibinfo
  {author} {\bibfnamefont {J.}~\bibnamefont {Marshall}}, \bibinfo {author} {\bibfnamefont {O.}~\bibnamefont {Martin}}, \bibinfo {author} {\bibfnamefont {J.~R.}\ \bibnamefont {McClean}}, \bibinfo {author} {\bibfnamefont {T.}~\bibnamefont {McCourt}}, \bibinfo {author} {\bibfnamefont {M.}~\bibnamefont {McEwen}}, \bibinfo {author} {\bibfnamefont {A.}~\bibnamefont {Megrant}}, \bibinfo {author} {\bibfnamefont {B.}~\bibnamefont {Meurer~Costa}}, \bibinfo {author} {\bibfnamefont {X.}~\bibnamefont {Mi}}, \bibinfo {author} {\bibfnamefont {K.~C.}\ \bibnamefont {Miao}}, \bibinfo {author} {\bibfnamefont {M.}~\bibnamefont {Mohseni}}, \bibinfo {author} {\bibfnamefont {S.}~\bibnamefont {Montazeri}}, \bibinfo {author} {\bibfnamefont {A.}~\bibnamefont {Morvan}}, \bibinfo {author} {\bibfnamefont {E.}~\bibnamefont {Mount}}, \bibinfo {author} {\bibfnamefont {W.}~\bibnamefont {Mruczkiewicz}}, \bibinfo {author} {\bibfnamefont {O.}~\bibnamefont {Naaman}}, \bibinfo {author} {\bibfnamefont {M.}~\bibnamefont {Neeley}}, \bibinfo {author}
  {\bibfnamefont {C.}~\bibnamefont {Neill}}, \bibinfo {author} {\bibfnamefont {A.}~\bibnamefont {Nersisyan}}, \bibinfo {author} {\bibfnamefont {H.}~\bibnamefont {Neven}}, \bibinfo {author} {\bibfnamefont {M.}~\bibnamefont {Newman}}, \bibinfo {author} {\bibfnamefont {J.~H.}\ \bibnamefont {Ng}}, \bibinfo {author} {\bibfnamefont {A.}~\bibnamefont {Nguyen}}, \bibinfo {author} {\bibfnamefont {M.}~\bibnamefont {Nguyen}}, \bibinfo {author} {\bibfnamefont {M.~Y.}\ \bibnamefont {Niu}}, \bibinfo {author} {\bibfnamefont {T.~E.}\ \bibnamefont {O'Brien}}, \bibinfo {author} {\bibfnamefont {A.}~\bibnamefont {Opremcak}}, \bibinfo {author} {\bibfnamefont {J.}~\bibnamefont {Platt}}, \bibinfo {author} {\bibfnamefont {A.}~\bibnamefont {Petukhov}}, \bibinfo {author} {\bibfnamefont {R.}~\bibnamefont {Potter}}, \bibinfo {author} {\bibfnamefont {L.~P.}\ \bibnamefont {Pryadko}}, \bibinfo {author} {\bibfnamefont {C.}~\bibnamefont {Quintana}}, \bibinfo {author} {\bibfnamefont {P.}~\bibnamefont {Roushan}}, \bibinfo {author}
  {\bibfnamefont {N.~C.}\ \bibnamefont {Rubin}}, \bibinfo {author} {\bibfnamefont {N.}~\bibnamefont {Saei}}, \bibinfo {author} {\bibfnamefont {D.}~\bibnamefont {Sank}}, \bibinfo {author} {\bibfnamefont {K.}~\bibnamefont {Sankaragomathi}}, \bibinfo {author} {\bibfnamefont {K.~J.}\ \bibnamefont {Satzinger}}, \bibinfo {author} {\bibfnamefont {H.~F.}\ \bibnamefont {Schurkus}}, \bibinfo {author} {\bibfnamefont {C.}~\bibnamefont {Schuster}}, \bibinfo {author} {\bibfnamefont {M.~J.}\ \bibnamefont {Shearn}}, \bibinfo {author} {\bibfnamefont {A.}~\bibnamefont {Shorter}}, \bibinfo {author} {\bibfnamefont {V.}~\bibnamefont {Shvarts}}, \bibinfo {author} {\bibfnamefont {J.}~\bibnamefont {Skruzny}}, \bibinfo {author} {\bibfnamefont {V.}~\bibnamefont {Smelyanskiy}}, \bibinfo {author} {\bibfnamefont {W.~C.}\ \bibnamefont {Smith}}, \bibinfo {author} {\bibfnamefont {G.}~\bibnamefont {Sterling}}, \bibinfo {author} {\bibfnamefont {D.}~\bibnamefont {Strain}}, \bibinfo {author} {\bibfnamefont {M.}~\bibnamefont {Szalay}}, \bibinfo
  {author} {\bibfnamefont {A.}~\bibnamefont {Torres}}, \bibinfo {author} {\bibfnamefont {G.}~\bibnamefont {Vidal}}, \bibinfo {author} {\bibfnamefont {B.}~\bibnamefont {Villalonga}}, \bibinfo {author} {\bibfnamefont {C.}~\bibnamefont {Vollgraff~Heidweiller}}, \bibinfo {author} {\bibfnamefont {T.}~\bibnamefont {White}}, \bibinfo {author} {\bibfnamefont {C.}~\bibnamefont {Xing}}, \bibinfo {author} {\bibfnamefont {Z.~J.}\ \bibnamefont {Yao}}, \bibinfo {author} {\bibfnamefont {P.}~\bibnamefont {Yeh}}, \bibinfo {author} {\bibfnamefont {J.}~\bibnamefont {Yoo}}, \bibinfo {author} {\bibfnamefont {G.}~\bibnamefont {Young}}, \bibinfo {author} {\bibfnamefont {A.}~\bibnamefont {Zalcman}}, \bibinfo {author} {\bibfnamefont {Y.}~\bibnamefont {Zhang}}, \bibinfo {author} {\bibfnamefont {N.}~\bibnamefont {Zhu}},\ and\ \bibinfo {author} {\bibfnamefont {G.~Q.}\ \bibnamefont {AI}},\ }\bibfield  {title} {\bibinfo {title} {Suppressing quantum errors by scaling a surface code logical qubit},\ }\href
  {https://doi.org/10.1038/s41586-022-05434-1} {\bibfield  {journal} {\bibinfo  {journal} {Nature}\ }\textbf {\bibinfo {volume} {614}},\ \bibinfo {pages} {676} (\bibinfo {year} {2023})}\BibitemShut {NoStop}%
\bibitem [{\citenamefont {Chen}\ \emph {et~al.}(2014)\citenamefont {Chen}, \citenamefont {Megrant}, \citenamefont {Kelly}, \citenamefont {Barends}, \citenamefont {Bochmann}, \citenamefont {Chen}, \citenamefont {Chiaro}, \citenamefont {Dunsworth}, \citenamefont {Jeffrey}, \citenamefont {Mutus}, \citenamefont {O'Malley}, \citenamefont {Neill}, \citenamefont {Roushan}, \citenamefont {Sank}, \citenamefont {Vainsencher}, \citenamefont {Wenner}, \citenamefont {White}, \citenamefont {Cleland},\ and\ \citenamefont {Martinis}}]{Chen2014}%
  \BibitemOpen
  \bibfield  {author} {\bibinfo {author} {\bibfnamefont {Z.}~\bibnamefont {Chen}}, \bibinfo {author} {\bibfnamefont {A.}~\bibnamefont {Megrant}}, \bibinfo {author} {\bibfnamefont {J.}~\bibnamefont {Kelly}}, \bibinfo {author} {\bibfnamefont {R.}~\bibnamefont {Barends}}, \bibinfo {author} {\bibfnamefont {J.}~\bibnamefont {Bochmann}}, \bibinfo {author} {\bibfnamefont {Y.}~\bibnamefont {Chen}}, \bibinfo {author} {\bibfnamefont {B.}~\bibnamefont {Chiaro}}, \bibinfo {author} {\bibfnamefont {A.}~\bibnamefont {Dunsworth}}, \bibinfo {author} {\bibfnamefont {E.}~\bibnamefont {Jeffrey}}, \bibinfo {author} {\bibfnamefont {J.~Y.}\ \bibnamefont {Mutus}}, \bibinfo {author} {\bibfnamefont {P.~J.~J.}\ \bibnamefont {O'Malley}}, \bibinfo {author} {\bibfnamefont {C.}~\bibnamefont {Neill}}, \bibinfo {author} {\bibfnamefont {P.}~\bibnamefont {Roushan}}, \bibinfo {author} {\bibfnamefont {D.}~\bibnamefont {Sank}}, \bibinfo {author} {\bibfnamefont {A.}~\bibnamefont {Vainsencher}}, \bibinfo {author} {\bibfnamefont {J.}~\bibnamefont
  {Wenner}}, \bibinfo {author} {\bibfnamefont {T.~C.}\ \bibnamefont {White}}, \bibinfo {author} {\bibfnamefont {A.~N.}\ \bibnamefont {Cleland}},\ and\ \bibinfo {author} {\bibfnamefont {J.~M.}\ \bibnamefont {Martinis}},\ }\bibfield  {title} {\bibinfo {title} {Fabrication and characterization of aluminum airbridges for superconducting microwave circuits},\ }\href {https://doi.org/10.1063/1.4863745} {\bibfield  {journal} {\bibinfo  {journal} {Applied Physics Letters}\ }\textbf {\bibinfo {volume} {104}},\ \bibinfo {pages} {052602} (\bibinfo {year} {2014})}\BibitemShut {NoStop}%
\bibitem [{\citenamefont {Yost}\ \emph {et~al.}(2020)\citenamefont {Yost}, \citenamefont {Schwartz}, \citenamefont {Mallek}, \citenamefont {Rosenberg}, \citenamefont {Stull}, \citenamefont {Yoder}, \citenamefont {Calusine}, \citenamefont {Cook}, \citenamefont {Das}, \citenamefont {Day}, \citenamefont {Golden}, \citenamefont {Kim}, \citenamefont {Melville}, \citenamefont {Niedzielski}, \citenamefont {Woods}, \citenamefont {Kerman},\ and\ \citenamefont {Oliver}}]{Yost2020}%
  \BibitemOpen
  \bibfield  {author} {\bibinfo {author} {\bibfnamefont {D.~R.~W.}\ \bibnamefont {Yost}}, \bibinfo {author} {\bibfnamefont {M.~E.}\ \bibnamefont {Schwartz}}, \bibinfo {author} {\bibfnamefont {J.}~\bibnamefont {Mallek}}, \bibinfo {author} {\bibfnamefont {D.}~\bibnamefont {Rosenberg}}, \bibinfo {author} {\bibfnamefont {C.}~\bibnamefont {Stull}}, \bibinfo {author} {\bibfnamefont {J.~L.}\ \bibnamefont {Yoder}}, \bibinfo {author} {\bibfnamefont {G.}~\bibnamefont {Calusine}}, \bibinfo {author} {\bibfnamefont {M.}~\bibnamefont {Cook}}, \bibinfo {author} {\bibfnamefont {R.}~\bibnamefont {Das}}, \bibinfo {author} {\bibfnamefont {A.~L.}\ \bibnamefont {Day}}, \bibinfo {author} {\bibfnamefont {E.~B.}\ \bibnamefont {Golden}}, \bibinfo {author} {\bibfnamefont {D.~K.}\ \bibnamefont {Kim}}, \bibinfo {author} {\bibfnamefont {A.}~\bibnamefont {Melville}}, \bibinfo {author} {\bibfnamefont {B.~M.}\ \bibnamefont {Niedzielski}}, \bibinfo {author} {\bibfnamefont {W.}~\bibnamefont {Woods}}, \bibinfo {author} {\bibfnamefont {A.~J.}\
  \bibnamefont {Kerman}},\ and\ \bibinfo {author} {\bibfnamefont {W.~D.}\ \bibnamefont {Oliver}},\ }\bibfield  {title} {\bibinfo {title} {Solid-state qubits integrated with superconducting through-silicon vias},\ }\href {https://doi.org/10.1038/s41534-020-00289-8} {\bibfield  {journal} {\bibinfo  {journal} {npj Quantum Information}\ }\textbf {\bibinfo {volume} {6}},\ \bibinfo {pages} {59} (\bibinfo {year} {2020})}\BibitemShut {NoStop}%
\bibitem [{\citenamefont {Vahidpour}\ \emph {et~al.}(2017)\citenamefont {Vahidpour}, \citenamefont {O'Brien}, \citenamefont {Whyland}, \citenamefont {Angeles}, \citenamefont {Marshall}, \citenamefont {Scarabelli}, \citenamefont {Crossman}, \citenamefont {Yadav}, \citenamefont {Mohan}, \citenamefont {Bui}, \citenamefont {Rawat}, \citenamefont {Renzas}, \citenamefont {Vodrahalli}, \citenamefont {Bestwick},\ and\ \citenamefont {Rigetti}}]{Vahidpour2017}%
  \BibitemOpen
  \bibfield  {author} {\bibinfo {author} {\bibfnamefont {M.}~\bibnamefont {Vahidpour}}, \bibinfo {author} {\bibfnamefont {W.}~\bibnamefont {O'Brien}}, \bibinfo {author} {\bibfnamefont {J.~T.}\ \bibnamefont {Whyland}}, \bibinfo {author} {\bibfnamefont {J.}~\bibnamefont {Angeles}}, \bibinfo {author} {\bibfnamefont {J.}~\bibnamefont {Marshall}}, \bibinfo {author} {\bibfnamefont {D.}~\bibnamefont {Scarabelli}}, \bibinfo {author} {\bibfnamefont {G.}~\bibnamefont {Crossman}}, \bibinfo {author} {\bibfnamefont {K.}~\bibnamefont {Yadav}}, \bibinfo {author} {\bibfnamefont {Y.}~\bibnamefont {Mohan}}, \bibinfo {author} {\bibfnamefont {C.}~\bibnamefont {Bui}}, \bibinfo {author} {\bibfnamefont {V.}~\bibnamefont {Rawat}}, \bibinfo {author} {\bibfnamefont {R.}~\bibnamefont {Renzas}}, \bibinfo {author} {\bibfnamefont {N.}~\bibnamefont {Vodrahalli}}, \bibinfo {author} {\bibfnamefont {A.}~\bibnamefont {Bestwick}},\ and\ \bibinfo {author} {\bibfnamefont {C.}~\bibnamefont {Rigetti}},\ }\bibfield  {title} {\bibinfo {title}
  {Superconducting through-silicon vias for quantum integrated circuits},\ }\href@noop {} {\bibfield  {journal} {\bibinfo  {journal} {arXiv preprint arXiv:1708.02226}\ } (\bibinfo {year} {2017})}\BibitemShut {NoStop}%
\bibitem [{\citenamefont {Alfaro-Barrantes}\ \emph {et~al.}(2020)\citenamefont {Alfaro-Barrantes}, \citenamefont {Mastrangeli}, \citenamefont {Thoen}, \citenamefont {Visser}, \citenamefont {Bueno}, \citenamefont {Baselmans},\ and\ \citenamefont {Sarro}}]{AlfaroBarrantes2020}%
  \BibitemOpen
  \bibfield  {author} {\bibinfo {author} {\bibfnamefont {J.}~\bibnamefont {Alfaro-Barrantes}}, \bibinfo {author} {\bibfnamefont {M.}~\bibnamefont {Mastrangeli}}, \bibinfo {author} {\bibfnamefont {D.}~\bibnamefont {Thoen}}, \bibinfo {author} {\bibfnamefont {S.}~\bibnamefont {Visser}}, \bibinfo {author} {\bibfnamefont {J.}~\bibnamefont {Bueno}}, \bibinfo {author} {\bibfnamefont {J.}~\bibnamefont {Baselmans}},\ and\ \bibinfo {author} {\bibfnamefont {P.}~\bibnamefont {Sarro}},\ }\bibfield  {title} {\bibinfo {title} {Superconducting high-aspect ratio through-silicon vias with dc-sputtered al for quantum 3d integration},\ }\href@noop {} {\bibfield  {journal} {\bibinfo  {journal} {IEEE Electron Device Letters}\ } (\bibinfo {year} {2020})}\BibitemShut {NoStop}%
\bibitem [{\citenamefont {Murray}\ and\ \citenamefont {Abraham}(2016)}]{Murray2016}%
  \BibitemOpen
  \bibfield  {author} {\bibinfo {author} {\bibfnamefont {C.~E.}\ \bibnamefont {Murray}}\ and\ \bibinfo {author} {\bibfnamefont {D.~W.}\ \bibnamefont {Abraham}},\ }\bibfield  {title} {\bibinfo {title} {Predicting substrate resonance mode frequency shifts using conductive, through-substrate vias},\ }\href@noop {} {\bibfield  {journal} {\bibinfo  {journal} {Applied Physics Letters}\ }\textbf {\bibinfo {volume} {108}},\ \bibinfo {pages} {084101} (\bibinfo {year} {2016})}\BibitemShut {NoStop}%
\bibitem [{\citenamefont {Spring}\ \emph {et~al.}(2020)\citenamefont {Spring}, \citenamefont {Tsunoda}, \citenamefont {Vlastakis},\ and\ \citenamefont {Leek}}]{Spring2020}%
  \BibitemOpen
  \bibfield  {author} {\bibinfo {author} {\bibfnamefont {P.~A.}\ \bibnamefont {Spring}}, \bibinfo {author} {\bibfnamefont {T.}~\bibnamefont {Tsunoda}}, \bibinfo {author} {\bibfnamefont {B.}~\bibnamefont {Vlastakis}},\ and\ \bibinfo {author} {\bibfnamefont {P.~J.}\ \bibnamefont {Leek}},\ }\bibfield  {title} {\bibinfo {title} {Modeling enclosures for large-scale superconducting quantum circuits},\ }\href@noop {} {\bibfield  {journal} {\bibinfo  {journal} {Physical Review Applied}\ }\textbf {\bibinfo {volume} {14}},\ \bibinfo {pages} {024061} (\bibinfo {year} {2020})}\BibitemShut {NoStop}%
\bibitem [{\citenamefont {Spring}\ \emph {et~al.}(2022)\citenamefont {Spring}, \citenamefont {Cao}, \citenamefont {Tsunoda}, \citenamefont {Campanaro}, \citenamefont {Fasciati}, \citenamefont {Wills}, \citenamefont {Bakr}, \citenamefont {Chidambaram}, \citenamefont {Shteynas}, \citenamefont {Carpenter}, \citenamefont {Gow}, \citenamefont {Gates}, \citenamefont {Vlastakis},\ and\ \citenamefont {Leek}}]{Spring2022}%
  \BibitemOpen
  \bibfield  {author} {\bibinfo {author} {\bibfnamefont {P.~A.}\ \bibnamefont {Spring}}, \bibinfo {author} {\bibfnamefont {S.}~\bibnamefont {Cao}}, \bibinfo {author} {\bibfnamefont {T.}~\bibnamefont {Tsunoda}}, \bibinfo {author} {\bibfnamefont {G.}~\bibnamefont {Campanaro}}, \bibinfo {author} {\bibfnamefont {S.}~\bibnamefont {Fasciati}}, \bibinfo {author} {\bibfnamefont {J.}~\bibnamefont {Wills}}, \bibinfo {author} {\bibfnamefont {M.}~\bibnamefont {Bakr}}, \bibinfo {author} {\bibfnamefont {V.}~\bibnamefont {Chidambaram}}, \bibinfo {author} {\bibfnamefont {B.}~\bibnamefont {Shteynas}}, \bibinfo {author} {\bibfnamefont {L.}~\bibnamefont {Carpenter}}, \bibinfo {author} {\bibfnamefont {P.}~\bibnamefont {Gow}}, \bibinfo {author} {\bibfnamefont {J.}~\bibnamefont {Gates}}, \bibinfo {author} {\bibfnamefont {B.}~\bibnamefont {Vlastakis}},\ and\ \bibinfo {author} {\bibfnamefont {P.~J.}\ \bibnamefont {Leek}},\ }\bibfield  {title} {\bibinfo {title} {High coherence and low cross-talk in a tileable 3d integrated
  superconducting circuit architecture},\ }\href {https://doi.org/10.1126/sciadv.abl6698} {\bibfield  {journal} {\bibinfo  {journal} {Science Advances}\ }\textbf {\bibinfo {volume} {8}},\ \bibinfo {pages} {eabl6698} (\bibinfo {year} {2022})}\BibitemShut {NoStop}%
\bibitem [{\citenamefont {Biznárová}\ \emph {et~al.}(2023)\citenamefont {Biznárová}, \citenamefont {Osman}, \citenamefont {Rehnman}, \citenamefont {Chayanun}, \citenamefont {Križan}, \citenamefont {Malmberg}, \citenamefont {Rommel}, \citenamefont {Warren}, \citenamefont {Delsing}, \citenamefont {Yurgens}, \citenamefont {Bylander},\ and\ \citenamefont {Roudsari}}]{Biznarova2023}%
  \BibitemOpen
  \bibfield  {author} {\bibinfo {author} {\bibfnamefont {J.}~\bibnamefont {Biznárová}}, \bibinfo {author} {\bibfnamefont {A.}~\bibnamefont {Osman}}, \bibinfo {author} {\bibfnamefont {E.}~\bibnamefont {Rehnman}}, \bibinfo {author} {\bibfnamefont {L.}~\bibnamefont {Chayanun}}, \bibinfo {author} {\bibfnamefont {C.}~\bibnamefont {Križan}}, \bibinfo {author} {\bibfnamefont {P.}~\bibnamefont {Malmberg}}, \bibinfo {author} {\bibfnamefont {M.}~\bibnamefont {Rommel}}, \bibinfo {author} {\bibfnamefont {C.}~\bibnamefont {Warren}}, \bibinfo {author} {\bibfnamefont {P.}~\bibnamefont {Delsing}}, \bibinfo {author} {\bibfnamefont {A.}~\bibnamefont {Yurgens}}, \bibinfo {author} {\bibfnamefont {J.}~\bibnamefont {Bylander}},\ and\ \bibinfo {author} {\bibfnamefont {A.~F.}\ \bibnamefont {Roudsari}},\ }\href@noop {} {\bibinfo {title} {Mitigation of interfacial dielectric loss in aluminum-on-silicon superconducting qubits}} (\bibinfo {year} {2023}),\ \Eprint {https://arxiv.org/abs/2310.06797} {arXiv:2310.06797 [quant-ph]}
  \BibitemShut {NoStop}%
\bibitem [{\citenamefont {Osman}\ \emph {et~al.}(2023)\citenamefont {Osman}, \citenamefont {Fern\'andez-Pend\'as}, \citenamefont {Warren}, \citenamefont {Kosen}, \citenamefont {Scigliuzzo}, \citenamefont {Frisk~Kockum}, \citenamefont {Tancredi}, \citenamefont {Fadavi~Roudsari},\ and\ \citenamefont {Bylander}}]{Osman2023}%
  \BibitemOpen
  \bibfield  {author} {\bibinfo {author} {\bibfnamefont {A.}~\bibnamefont {Osman}}, \bibinfo {author} {\bibfnamefont {J.}~\bibnamefont {Fern\'andez-Pend\'as}}, \bibinfo {author} {\bibfnamefont {C.}~\bibnamefont {Warren}}, \bibinfo {author} {\bibfnamefont {S.}~\bibnamefont {Kosen}}, \bibinfo {author} {\bibfnamefont {M.}~\bibnamefont {Scigliuzzo}}, \bibinfo {author} {\bibfnamefont {A.}~\bibnamefont {Frisk~Kockum}}, \bibinfo {author} {\bibfnamefont {G.}~\bibnamefont {Tancredi}}, \bibinfo {author} {\bibfnamefont {A.}~\bibnamefont {Fadavi~Roudsari}},\ and\ \bibinfo {author} {\bibfnamefont {J.}~\bibnamefont {Bylander}},\ }\bibfield  {title} {\bibinfo {title} {Mitigation of frequency collisions in superconducting quantum processors},\ }\href {https://doi.org/10.1103/PhysRevResearch.5.043001} {\bibfield  {journal} {\bibinfo  {journal} {Phys. Rev. Res.}\ }\textbf {\bibinfo {volume} {5}},\ \bibinfo {pages} {043001} (\bibinfo {year} {2023})}\BibitemShut {NoStop}%
\bibitem [{\citenamefont {Bal}\ \emph {et~al.}(2024)\citenamefont {Bal}, \citenamefont {Murthy}, \citenamefont {Zhu}, \citenamefont {Crisa}, \citenamefont {You}, \citenamefont {Huang}, \citenamefont {Roy}, \citenamefont {Lee}, \citenamefont {Zanten}, \citenamefont {Pilipenko}, \citenamefont {Nekrashevich}, \citenamefont {Lunin}, \citenamefont {Bafia}, \citenamefont {Krasnikova}, \citenamefont {Kopas}, \citenamefont {Lachman}, \citenamefont {Miller}, \citenamefont {Mutus}, \citenamefont {Reagor}, \citenamefont {Cansizoglu}, \citenamefont {Marshall}, \citenamefont {Pappas}, \citenamefont {Vu}, \citenamefont {Yadavalli}, \citenamefont {Oh}, \citenamefont {Zhou}, \citenamefont {Kramer}, \citenamefont {Lecocq}, \citenamefont {Goronzy}, \citenamefont {Torres-Castanedo}, \citenamefont {Pritchard}, \citenamefont {Dravid}, \citenamefont {Rondinelli}, \citenamefont {Bedzyk}, \citenamefont {Hersam}, \citenamefont {Zasadzinski}, \citenamefont {Koch}, \citenamefont {Sauls}, \citenamefont {Romanenko},\ and\ \citenamefont
  {Grassellino}}]{Bal2024}%
  \BibitemOpen
  \bibfield  {author} {\bibinfo {author} {\bibfnamefont {M.}~\bibnamefont {Bal}}, \bibinfo {author} {\bibfnamefont {A.~A.}\ \bibnamefont {Murthy}}, \bibinfo {author} {\bibfnamefont {S.}~\bibnamefont {Zhu}}, \bibinfo {author} {\bibfnamefont {F.}~\bibnamefont {Crisa}}, \bibinfo {author} {\bibfnamefont {X.}~\bibnamefont {You}}, \bibinfo {author} {\bibfnamefont {Z.}~\bibnamefont {Huang}}, \bibinfo {author} {\bibfnamefont {T.}~\bibnamefont {Roy}}, \bibinfo {author} {\bibfnamefont {J.}~\bibnamefont {Lee}}, \bibinfo {author} {\bibfnamefont {D.~v.}\ \bibnamefont {Zanten}}, \bibinfo {author} {\bibfnamefont {R.}~\bibnamefont {Pilipenko}}, \bibinfo {author} {\bibfnamefont {I.}~\bibnamefont {Nekrashevich}}, \bibinfo {author} {\bibfnamefont {A.}~\bibnamefont {Lunin}}, \bibinfo {author} {\bibfnamefont {D.}~\bibnamefont {Bafia}}, \bibinfo {author} {\bibfnamefont {Y.}~\bibnamefont {Krasnikova}}, \bibinfo {author} {\bibfnamefont {C.~J.}\ \bibnamefont {Kopas}}, \bibinfo {author} {\bibfnamefont {E.~O.}\ \bibnamefont {Lachman}},
  \bibinfo {author} {\bibfnamefont {D.}~\bibnamefont {Miller}}, \bibinfo {author} {\bibfnamefont {J.~Y.}\ \bibnamefont {Mutus}}, \bibinfo {author} {\bibfnamefont {M.~J.}\ \bibnamefont {Reagor}}, \bibinfo {author} {\bibfnamefont {H.}~\bibnamefont {Cansizoglu}}, \bibinfo {author} {\bibfnamefont {J.}~\bibnamefont {Marshall}}, \bibinfo {author} {\bibfnamefont {D.~P.}\ \bibnamefont {Pappas}}, \bibinfo {author} {\bibfnamefont {K.}~\bibnamefont {Vu}}, \bibinfo {author} {\bibfnamefont {K.}~\bibnamefont {Yadavalli}}, \bibinfo {author} {\bibfnamefont {J.-S.}\ \bibnamefont {Oh}}, \bibinfo {author} {\bibfnamefont {L.}~\bibnamefont {Zhou}}, \bibinfo {author} {\bibfnamefont {M.~J.}\ \bibnamefont {Kramer}}, \bibinfo {author} {\bibfnamefont {F.}~\bibnamefont {Lecocq}}, \bibinfo {author} {\bibfnamefont {D.~P.}\ \bibnamefont {Goronzy}}, \bibinfo {author} {\bibfnamefont {C.~G.}\ \bibnamefont {Torres-Castanedo}}, \bibinfo {author} {\bibfnamefont {P.~G.}\ \bibnamefont {Pritchard}}, \bibinfo {author} {\bibfnamefont {V.~P.}\
  \bibnamefont {Dravid}}, \bibinfo {author} {\bibfnamefont {J.~M.}\ \bibnamefont {Rondinelli}}, \bibinfo {author} {\bibfnamefont {M.~J.}\ \bibnamefont {Bedzyk}}, \bibinfo {author} {\bibfnamefont {M.~C.}\ \bibnamefont {Hersam}}, \bibinfo {author} {\bibfnamefont {J.}~\bibnamefont {Zasadzinski}}, \bibinfo {author} {\bibfnamefont {J.}~\bibnamefont {Koch}}, \bibinfo {author} {\bibfnamefont {J.~A.}\ \bibnamefont {Sauls}}, \bibinfo {author} {\bibfnamefont {A.}~\bibnamefont {Romanenko}},\ and\ \bibinfo {author} {\bibfnamefont {A.}~\bibnamefont {Grassellino}},\ }\bibfield  {title} {\bibinfo {title} {Systematic improvements in transmon qubit coherence enabled by niobium surface encapsulation},\ }\href {https://doi.org/10.1038/s41534-024-00840-x} {\bibfield  {journal} {\bibinfo  {journal} {npj Quantum Information}\ }\textbf {\bibinfo {volume} {10}},\ \bibinfo {pages} {43} (\bibinfo {year} {2024})}\BibitemShut {NoStop}%
\bibitem [{\citenamefont {Place}\ \emph {et~al.}(2021)\citenamefont {Place}, \citenamefont {Rodgers}, \citenamefont {Mundada}, \citenamefont {Smitham}, \citenamefont {Fitzpatrick}, \citenamefont {Leng}, \citenamefont {Premkumar}, \citenamefont {Bryon}, \citenamefont {Vrajitoarea}, \citenamefont {Sussman}, \citenamefont {Cheng}, \citenamefont {Madhavan}, \citenamefont {Babla}, \citenamefont {Le}, \citenamefont {Gang}, \citenamefont {J{\"a}ck}, \citenamefont {Gyenis}, \citenamefont {Yao}, \citenamefont {Cava}, \citenamefont {de~Leon},\ and\ \citenamefont {Houck}}]{Place2021}%
  \BibitemOpen
  \bibfield  {author} {\bibinfo {author} {\bibfnamefont {A.~P.~M.}\ \bibnamefont {Place}}, \bibinfo {author} {\bibfnamefont {L.~V.~H.}\ \bibnamefont {Rodgers}}, \bibinfo {author} {\bibfnamefont {P.}~\bibnamefont {Mundada}}, \bibinfo {author} {\bibfnamefont {B.~M.}\ \bibnamefont {Smitham}}, \bibinfo {author} {\bibfnamefont {M.}~\bibnamefont {Fitzpatrick}}, \bibinfo {author} {\bibfnamefont {Z.}~\bibnamefont {Leng}}, \bibinfo {author} {\bibfnamefont {A.}~\bibnamefont {Premkumar}}, \bibinfo {author} {\bibfnamefont {J.}~\bibnamefont {Bryon}}, \bibinfo {author} {\bibfnamefont {A.}~\bibnamefont {Vrajitoarea}}, \bibinfo {author} {\bibfnamefont {S.}~\bibnamefont {Sussman}}, \bibinfo {author} {\bibfnamefont {G.}~\bibnamefont {Cheng}}, \bibinfo {author} {\bibfnamefont {T.}~\bibnamefont {Madhavan}}, \bibinfo {author} {\bibfnamefont {H.~K.}\ \bibnamefont {Babla}}, \bibinfo {author} {\bibfnamefont {X.~H.}\ \bibnamefont {Le}}, \bibinfo {author} {\bibfnamefont {Y.}~\bibnamefont {Gang}}, \bibinfo {author} {\bibfnamefont
  {B.}~\bibnamefont {J{\"a}ck}}, \bibinfo {author} {\bibfnamefont {A.}~\bibnamefont {Gyenis}}, \bibinfo {author} {\bibfnamefont {N.}~\bibnamefont {Yao}}, \bibinfo {author} {\bibfnamefont {R.~J.}\ \bibnamefont {Cava}}, \bibinfo {author} {\bibfnamefont {N.~P.}\ \bibnamefont {de~Leon}},\ and\ \bibinfo {author} {\bibfnamefont {A.~A.}\ \bibnamefont {Houck}},\ }\bibfield  {title} {\bibinfo {title} {New material platform for superconducting transmon qubits with coherence times exceeding 0.3 milliseconds},\ }\href {https://doi.org/10.1038/s41467-021-22030-5} {\bibfield  {journal} {\bibinfo  {journal} {Nature Communications}\ }\textbf {\bibinfo {volume} {12}},\ \bibinfo {pages} {1779} (\bibinfo {year} {2021})}\BibitemShut {NoStop}%
\bibitem [{\citenamefont {Wang}\ \emph {et~al.}(2022)\citenamefont {Wang}, \citenamefont {Li}, \citenamefont {Xu}, \citenamefont {Li}, \citenamefont {Wang}, \citenamefont {Yang}, \citenamefont {Mi}, \citenamefont {Liang}, \citenamefont {Su}, \citenamefont {Yang}, \citenamefont {Wang}, \citenamefont {Wang}, \citenamefont {Li}, \citenamefont {Chen}, \citenamefont {Li}, \citenamefont {Linghu}, \citenamefont {Han}, \citenamefont {Zhang}, \citenamefont {Feng}, \citenamefont {Song}, \citenamefont {Ma}, \citenamefont {Zhang}, \citenamefont {Wang}, \citenamefont {Zhao}, \citenamefont {Liu}, \citenamefont {Xue}, \citenamefont {Jin},\ and\ \citenamefont {Yu}}]{Wang2022}%
  \BibitemOpen
  \bibfield  {author} {\bibinfo {author} {\bibfnamefont {C.}~\bibnamefont {Wang}}, \bibinfo {author} {\bibfnamefont {X.}~\bibnamefont {Li}}, \bibinfo {author} {\bibfnamefont {H.}~\bibnamefont {Xu}}, \bibinfo {author} {\bibfnamefont {Z.}~\bibnamefont {Li}}, \bibinfo {author} {\bibfnamefont {J.}~\bibnamefont {Wang}}, \bibinfo {author} {\bibfnamefont {Z.}~\bibnamefont {Yang}}, \bibinfo {author} {\bibfnamefont {Z.}~\bibnamefont {Mi}}, \bibinfo {author} {\bibfnamefont {X.}~\bibnamefont {Liang}}, \bibinfo {author} {\bibfnamefont {T.}~\bibnamefont {Su}}, \bibinfo {author} {\bibfnamefont {C.}~\bibnamefont {Yang}}, \bibinfo {author} {\bibfnamefont {G.}~\bibnamefont {Wang}}, \bibinfo {author} {\bibfnamefont {W.}~\bibnamefont {Wang}}, \bibinfo {author} {\bibfnamefont {Y.}~\bibnamefont {Li}}, \bibinfo {author} {\bibfnamefont {M.}~\bibnamefont {Chen}}, \bibinfo {author} {\bibfnamefont {C.}~\bibnamefont {Li}}, \bibinfo {author} {\bibfnamefont {K.}~\bibnamefont {Linghu}}, \bibinfo {author} {\bibfnamefont {J.}~\bibnamefont
  {Han}}, \bibinfo {author} {\bibfnamefont {Y.}~\bibnamefont {Zhang}}, \bibinfo {author} {\bibfnamefont {Y.}~\bibnamefont {Feng}}, \bibinfo {author} {\bibfnamefont {Y.}~\bibnamefont {Song}}, \bibinfo {author} {\bibfnamefont {T.}~\bibnamefont {Ma}}, \bibinfo {author} {\bibfnamefont {J.}~\bibnamefont {Zhang}}, \bibinfo {author} {\bibfnamefont {R.}~\bibnamefont {Wang}}, \bibinfo {author} {\bibfnamefont {P.}~\bibnamefont {Zhao}}, \bibinfo {author} {\bibfnamefont {W.}~\bibnamefont {Liu}}, \bibinfo {author} {\bibfnamefont {G.}~\bibnamefont {Xue}}, \bibinfo {author} {\bibfnamefont {Y.}~\bibnamefont {Jin}},\ and\ \bibinfo {author} {\bibfnamefont {H.}~\bibnamefont {Yu}},\ }\bibfield  {title} {\bibinfo {title} {Towards practical quantum computers: transmon qubit with a lifetime approaching 0.5 milliseconds},\ }\href {https://doi.org/10.1038/s41534-021-00510-2} {\bibfield  {journal} {\bibinfo  {journal} {npj Quantum Information}\ }\textbf {\bibinfo {volume} {8}},\ \bibinfo {pages} {3} (\bibinfo {year}
  {2022})}\BibitemShut {NoStop}%
\bibitem [{\citenamefont {Deng}\ \emph {et~al.}(2023)\citenamefont {Deng}, \citenamefont {Song}, \citenamefont {Gao}, \citenamefont {Xia}, \citenamefont {Bao}, \citenamefont {Jiang}, \citenamefont {Ku}, \citenamefont {Li}, \citenamefont {Ma}, \citenamefont {Qin}, \citenamefont {Sun}, \citenamefont {Tang}, \citenamefont {Wang}, \citenamefont {Wu}, \citenamefont {Yu}, \citenamefont {Zhang}, \citenamefont {Zhang}, \citenamefont {Zhou}, \citenamefont {Zhu}, \citenamefont {Shi}, \citenamefont {Zhao},\ and\ \citenamefont {Deng}}]{Deng2023}%
  \BibitemOpen
  \bibfield  {author} {\bibinfo {author} {\bibfnamefont {H.}~\bibnamefont {Deng}}, \bibinfo {author} {\bibfnamefont {Z.}~\bibnamefont {Song}}, \bibinfo {author} {\bibfnamefont {R.}~\bibnamefont {Gao}}, \bibinfo {author} {\bibfnamefont {T.}~\bibnamefont {Xia}}, \bibinfo {author} {\bibfnamefont {F.}~\bibnamefont {Bao}}, \bibinfo {author} {\bibfnamefont {X.}~\bibnamefont {Jiang}}, \bibinfo {author} {\bibfnamefont {H.-S.}\ \bibnamefont {Ku}}, \bibinfo {author} {\bibfnamefont {Z.}~\bibnamefont {Li}}, \bibinfo {author} {\bibfnamefont {X.}~\bibnamefont {Ma}}, \bibinfo {author} {\bibfnamefont {J.}~\bibnamefont {Qin}}, \bibinfo {author} {\bibfnamefont {H.}~\bibnamefont {Sun}}, \bibinfo {author} {\bibfnamefont {C.}~\bibnamefont {Tang}}, \bibinfo {author} {\bibfnamefont {T.}~\bibnamefont {Wang}}, \bibinfo {author} {\bibfnamefont {F.}~\bibnamefont {Wu}}, \bibinfo {author} {\bibfnamefont {W.}~\bibnamefont {Yu}}, \bibinfo {author} {\bibfnamefont {G.}~\bibnamefont {Zhang}}, \bibinfo {author} {\bibfnamefont {X.}~\bibnamefont
  {Zhang}}, \bibinfo {author} {\bibfnamefont {J.}~\bibnamefont {Zhou}}, \bibinfo {author} {\bibfnamefont {X.}~\bibnamefont {Zhu}}, \bibinfo {author} {\bibfnamefont {Y.}~\bibnamefont {Shi}}, \bibinfo {author} {\bibfnamefont {H.-H.}\ \bibnamefont {Zhao}},\ and\ \bibinfo {author} {\bibfnamefont {C.}~\bibnamefont {Deng}},\ }\bibfield  {title} {\bibinfo {title} {Titanium nitride film on sapphire substrate with low dielectric loss for superconducting qubits},\ }\href {https://doi.org/10.1103/PhysRevApplied.19.024013} {\bibfield  {journal} {\bibinfo  {journal} {Phys. Rev. Appl.}\ }\textbf {\bibinfo {volume} {19}},\ \bibinfo {pages} {024013} (\bibinfo {year} {2023})}\BibitemShut {NoStop}%
\bibitem [{\citenamefont {Kim}\ \emph {et~al.}(2021)\citenamefont {Kim}, \citenamefont {Terai}, \citenamefont {Yamashita}, \citenamefont {Qiu}, \citenamefont {Fuse}, \citenamefont {Yoshihara}, \citenamefont {Ashhab}, \citenamefont {Inomata},\ and\ \citenamefont {Semba}}]{Kim2021}%
  \BibitemOpen
  \bibfield  {author} {\bibinfo {author} {\bibfnamefont {S.}~\bibnamefont {Kim}}, \bibinfo {author} {\bibfnamefont {H.}~\bibnamefont {Terai}}, \bibinfo {author} {\bibfnamefont {T.}~\bibnamefont {Yamashita}}, \bibinfo {author} {\bibfnamefont {W.}~\bibnamefont {Qiu}}, \bibinfo {author} {\bibfnamefont {T.}~\bibnamefont {Fuse}}, \bibinfo {author} {\bibfnamefont {F.}~\bibnamefont {Yoshihara}}, \bibinfo {author} {\bibfnamefont {S.}~\bibnamefont {Ashhab}}, \bibinfo {author} {\bibfnamefont {K.}~\bibnamefont {Inomata}},\ and\ \bibinfo {author} {\bibfnamefont {K.}~\bibnamefont {Semba}},\ }\bibfield  {title} {\bibinfo {title} {Enhanced coherence of all-nitride superconducting qubits epitaxially grown on silicon substrate},\ }\href {https://doi.org/10.1038/s43246-021-00204-4} {\bibfield  {journal} {\bibinfo  {journal} {Communications Materials}\ }\textbf {\bibinfo {volume} {2}},\ \bibinfo {pages} {98} (\bibinfo {year} {2021})}\BibitemShut {NoStop}%
\bibitem [{\citenamefont {Hazard}\ \emph {et~al.}(2023)\citenamefont {Hazard}, \citenamefont {Woods}, \citenamefont {Rosenberg}, \citenamefont {Das}, \citenamefont {Hirjibehedin}, \citenamefont {Kim}, \citenamefont {Knecht}, \citenamefont {Mallek}, \citenamefont {Melville}, \citenamefont {Niedzielski}, \citenamefont {Serniak}, \citenamefont {Sliwa}, \citenamefont {Yost}, \citenamefont {Yoder}, \citenamefont {Oliver},\ and\ \citenamefont {Schwartz}}]{Hazard2023}%
  \BibitemOpen
  \bibfield  {author} {\bibinfo {author} {\bibfnamefont {T.~M.}\ \bibnamefont {Hazard}}, \bibinfo {author} {\bibfnamefont {W.}~\bibnamefont {Woods}}, \bibinfo {author} {\bibfnamefont {D.}~\bibnamefont {Rosenberg}}, \bibinfo {author} {\bibfnamefont {R.}~\bibnamefont {Das}}, \bibinfo {author} {\bibfnamefont {C.~F.}\ \bibnamefont {Hirjibehedin}}, \bibinfo {author} {\bibfnamefont {D.~K.}\ \bibnamefont {Kim}}, \bibinfo {author} {\bibfnamefont {J.~M.}\ \bibnamefont {Knecht}}, \bibinfo {author} {\bibfnamefont {J.}~\bibnamefont {Mallek}}, \bibinfo {author} {\bibfnamefont {A.}~\bibnamefont {Melville}}, \bibinfo {author} {\bibfnamefont {B.~M.}\ \bibnamefont {Niedzielski}}, \bibinfo {author} {\bibfnamefont {K.}~\bibnamefont {Serniak}}, \bibinfo {author} {\bibfnamefont {K.~M.}\ \bibnamefont {Sliwa}}, \bibinfo {author} {\bibfnamefont {D.~R.~W.}\ \bibnamefont {Yost}}, \bibinfo {author} {\bibfnamefont {J.~L.}\ \bibnamefont {Yoder}}, \bibinfo {author} {\bibfnamefont {W.~D.}\ \bibnamefont {Oliver}},\ and\ \bibinfo {author}
  {\bibfnamefont {M.~E.}\ \bibnamefont {Schwartz}},\ }\bibfield  {title} {\bibinfo {title} {{Characterization of superconducting through-silicon vias as capacitive elements in quantum circuits}},\ }\href@noop {} {\bibfield  {journal} {\bibinfo  {journal} {Applied Physics Letters}\ }\textbf {\bibinfo {volume} {123}},\ \bibinfo {pages} {154004} (\bibinfo {year} {2023})}\BibitemShut {NoStop}%
\bibitem [{\citenamefont {Kono}\ \emph {et~al.}(2024)\citenamefont {Kono}, \citenamefont {Pan}, \citenamefont {Chegnizadeh}, \citenamefont {Wang}, \citenamefont {Youssefi}, \citenamefont {Scigliuzzo},\ and\ \citenamefont {Kippenberg}}]{Kono2024}%
  \BibitemOpen
  \bibfield  {author} {\bibinfo {author} {\bibfnamefont {S.}~\bibnamefont {Kono}}, \bibinfo {author} {\bibfnamefont {J.}~\bibnamefont {Pan}}, \bibinfo {author} {\bibfnamefont {M.}~\bibnamefont {Chegnizadeh}}, \bibinfo {author} {\bibfnamefont {X.}~\bibnamefont {Wang}}, \bibinfo {author} {\bibfnamefont {A.}~\bibnamefont {Youssefi}}, \bibinfo {author} {\bibfnamefont {M.}~\bibnamefont {Scigliuzzo}},\ and\ \bibinfo {author} {\bibfnamefont {T.~J.}\ \bibnamefont {Kippenberg}},\ }\bibfield  {title} {\bibinfo {title} {Mechanically induced correlated errors on superconducting qubits with relaxation times exceeding 0.4{\thinspace}ms},\ }\href {https://doi.org/10.1038/s41467-024-48230-3} {\bibfield  {journal} {\bibinfo  {journal} {Nature Communications}\ }\textbf {\bibinfo {volume} {15}},\ \bibinfo {pages} {3950} (\bibinfo {year} {2024})}\BibitemShut {NoStop}%
\bibitem [{\citenamefont {Read}\ \emph {et~al.}(2023)\citenamefont {Read}, \citenamefont {Chapman}, \citenamefont {Lei}, \citenamefont {Curtis}, \citenamefont {Ganjam}, \citenamefont {Krayzman}, \citenamefont {Frunzio},\ and\ \citenamefont {Schoelkopf}}]{Read2023}%
  \BibitemOpen
  \bibfield  {author} {\bibinfo {author} {\bibfnamefont {A.~P.}\ \bibnamefont {Read}}, \bibinfo {author} {\bibfnamefont {B.~J.}\ \bibnamefont {Chapman}}, \bibinfo {author} {\bibfnamefont {C.~U.}\ \bibnamefont {Lei}}, \bibinfo {author} {\bibfnamefont {J.~C.}\ \bibnamefont {Curtis}}, \bibinfo {author} {\bibfnamefont {S.}~\bibnamefont {Ganjam}}, \bibinfo {author} {\bibfnamefont {L.}~\bibnamefont {Krayzman}}, \bibinfo {author} {\bibfnamefont {L.}~\bibnamefont {Frunzio}},\ and\ \bibinfo {author} {\bibfnamefont {R.~J.}\ \bibnamefont {Schoelkopf}},\ }\bibfield  {title} {\bibinfo {title} {Precision measurement of the microwave dielectric loss of sapphire in the quantum regime with parts-per-billion sensitivity},\ }\href {https://doi.org/10.1103/PhysRevApplied.19.034064} {\bibfield  {journal} {\bibinfo  {journal} {Phys. Rev. Appl.}\ }\textbf {\bibinfo {volume} {19}},\ \bibinfo {pages} {034064} (\bibinfo {year} {2023})}\BibitemShut {NoStop}%
\bibitem [{\citenamefont {Checchin}\ \emph {et~al.}(2022)\citenamefont {Checchin}, \citenamefont {Frolov}, \citenamefont {Lunin}, \citenamefont {Grassellino},\ and\ \citenamefont {Romanenko}}]{Checchin2022}%
  \BibitemOpen
  \bibfield  {author} {\bibinfo {author} {\bibfnamefont {M.}~\bibnamefont {Checchin}}, \bibinfo {author} {\bibfnamefont {D.}~\bibnamefont {Frolov}}, \bibinfo {author} {\bibfnamefont {A.}~\bibnamefont {Lunin}}, \bibinfo {author} {\bibfnamefont {A.}~\bibnamefont {Grassellino}},\ and\ \bibinfo {author} {\bibfnamefont {A.}~\bibnamefont {Romanenko}},\ }\bibfield  {title} {\bibinfo {title} {Measurement of the low-temperature loss tangent of high-resistivity silicon using a high-$q$ superconducting resonator},\ }\href {https://doi.org/10.1103/PhysRevApplied.18.034013} {\bibfield  {journal} {\bibinfo  {journal} {Phys. Rev. Appl.}\ }\textbf {\bibinfo {volume} {18}},\ \bibinfo {pages} {034013} (\bibinfo {year} {2022})}\BibitemShut {NoStop}%
\bibitem [{\citenamefont {Zhang}\ \emph {et~al.}(2024)\citenamefont {Zhang}, \citenamefont {Godeneli}, \citenamefont {He}, \citenamefont {Odeh}, \citenamefont {Zhou}, \citenamefont {Meesala},\ and\ \citenamefont {Sipahigil}}]{Zhang2024}%
  \BibitemOpen
  \bibfield  {author} {\bibinfo {author} {\bibfnamefont {Z.-H.}\ \bibnamefont {Zhang}}, \bibinfo {author} {\bibfnamefont {K.}~\bibnamefont {Godeneli}}, \bibinfo {author} {\bibfnamefont {J.}~\bibnamefont {He}}, \bibinfo {author} {\bibfnamefont {M.}~\bibnamefont {Odeh}}, \bibinfo {author} {\bibfnamefont {H.}~\bibnamefont {Zhou}}, \bibinfo {author} {\bibfnamefont {S.}~\bibnamefont {Meesala}},\ and\ \bibinfo {author} {\bibfnamefont {A.}~\bibnamefont {Sipahigil}},\ }\href@noop {} {\bibinfo {title} {Acceptor-induced bulk dielectric loss in superconducting circuits on silicon}} (\bibinfo {year} {2024}),\ \Eprint {https://arxiv.org/abs/2402.17155} {arXiv:2402.17155 [quant-ph]} \BibitemShut {NoStop}%
\bibitem [{\citenamefont {Rahamim}\ \emph {et~al.}(2017)\citenamefont {Rahamim}, \citenamefont {Behrle}, \citenamefont {Peterer}, \citenamefont {Patterson}, \citenamefont {Spring}, \citenamefont {Tsunoda}, \citenamefont {Manenti}, \citenamefont {Tancredi},\ and\ \citenamefont {Leek}}]{Rahamim2017}%
  \BibitemOpen
  \bibfield  {author} {\bibinfo {author} {\bibfnamefont {J.}~\bibnamefont {Rahamim}}, \bibinfo {author} {\bibfnamefont {T.}~\bibnamefont {Behrle}}, \bibinfo {author} {\bibfnamefont {M.}~\bibnamefont {Peterer}}, \bibinfo {author} {\bibfnamefont {A.~D.}\ \bibnamefont {Patterson}}, \bibinfo {author} {\bibfnamefont {P.~A.}\ \bibnamefont {Spring}}, \bibinfo {author} {\bibfnamefont {T.}~\bibnamefont {Tsunoda}}, \bibinfo {author} {\bibfnamefont {R.}~\bibnamefont {Manenti}}, \bibinfo {author} {\bibfnamefont {G.}~\bibnamefont {Tancredi}},\ and\ \bibinfo {author} {\bibfnamefont {P.~J.}\ \bibnamefont {Leek}},\ }\bibfield  {title} {\bibinfo {title} {Double-sided coaxial circuit qed with out-of-plane wiring},\ }\href@noop {} {\bibfield  {journal} {\bibinfo  {journal} {Applied Physics Letters}\ }\textbf {\bibinfo {volume} {110}},\ \bibinfo {pages} {222602} (\bibinfo {year} {2017})}\BibitemShut {NoStop}%
\bibitem [{\citenamefont {Wu}\ \emph {et~al.}(2010)\citenamefont {Wu}, \citenamefont {Kumar},\ and\ \citenamefont {Pamarthy}}]{Wu2010}%
  \BibitemOpen
  \bibfield  {author} {\bibinfo {author} {\bibfnamefont {B.}~\bibnamefont {Wu}}, \bibinfo {author} {\bibfnamefont {A.}~\bibnamefont {Kumar}},\ and\ \bibinfo {author} {\bibfnamefont {S.}~\bibnamefont {Pamarthy}},\ }\bibfield  {title} {\bibinfo {title} {{High aspect ratio silicon etch: A review}},\ }\href {https://doi.org/10.1063/1.3474652} {\bibfield  {journal} {\bibinfo  {journal} {Journal of Applied Physics}\ }\textbf {\bibinfo {volume} {108}},\ \bibinfo {pages} {051101} (\bibinfo {year} {2010})}\BibitemShut {NoStop}%
\bibitem [{\citenamefont {Grigoras}\ \emph {et~al.}(2022)\citenamefont {Grigoras}, \citenamefont {Yurttagül}, \citenamefont {Kaikkonen}, \citenamefont {Mannila}, \citenamefont {Eskelinen}, \citenamefont {Lozano}, \citenamefont {Li}, \citenamefont {Rommel}, \citenamefont {Shiri}, \citenamefont {Tiencken}, \citenamefont {Simbierowicz}, \citenamefont {Ronzani}, \citenamefont {Hätinen}, \citenamefont {Datta}, \citenamefont {Vesterinen}, \citenamefont {Grönberg}, \citenamefont {Biznárová}, \citenamefont {Roudsari}, \citenamefont {Kosen}, \citenamefont {Osman}, \citenamefont {Prunnila}, \citenamefont {Hassel}, \citenamefont {Bylander},\ and\ \citenamefont {Govenius}}]{Grigoras2022}%
  \BibitemOpen
  \bibfield  {author} {\bibinfo {author} {\bibfnamefont {K.}~\bibnamefont {Grigoras}}, \bibinfo {author} {\bibfnamefont {N.}~\bibnamefont {Yurttagül}}, \bibinfo {author} {\bibfnamefont {J.-P.}\ \bibnamefont {Kaikkonen}}, \bibinfo {author} {\bibfnamefont {E.~T.}\ \bibnamefont {Mannila}}, \bibinfo {author} {\bibfnamefont {P.}~\bibnamefont {Eskelinen}}, \bibinfo {author} {\bibfnamefont {D.~P.}\ \bibnamefont {Lozano}}, \bibinfo {author} {\bibfnamefont {H.-X.}\ \bibnamefont {Li}}, \bibinfo {author} {\bibfnamefont {M.}~\bibnamefont {Rommel}}, \bibinfo {author} {\bibfnamefont {D.}~\bibnamefont {Shiri}}, \bibinfo {author} {\bibfnamefont {N.}~\bibnamefont {Tiencken}}, \bibinfo {author} {\bibfnamefont {S.}~\bibnamefont {Simbierowicz}}, \bibinfo {author} {\bibfnamefont {A.}~\bibnamefont {Ronzani}}, \bibinfo {author} {\bibfnamefont {J.}~\bibnamefont {Hätinen}}, \bibinfo {author} {\bibfnamefont {D.}~\bibnamefont {Datta}}, \bibinfo {author} {\bibfnamefont {V.}~\bibnamefont {Vesterinen}}, \bibinfo {author} {\bibfnamefont
  {L.}~\bibnamefont {Grönberg}}, \bibinfo {author} {\bibfnamefont {J.}~\bibnamefont {Biznárová}}, \bibinfo {author} {\bibfnamefont {A.~F.}\ \bibnamefont {Roudsari}}, \bibinfo {author} {\bibfnamefont {S.}~\bibnamefont {Kosen}}, \bibinfo {author} {\bibfnamefont {A.}~\bibnamefont {Osman}}, \bibinfo {author} {\bibfnamefont {M.}~\bibnamefont {Prunnila}}, \bibinfo {author} {\bibfnamefont {J.}~\bibnamefont {Hassel}}, \bibinfo {author} {\bibfnamefont {J.}~\bibnamefont {Bylander}},\ and\ \bibinfo {author} {\bibfnamefont {J.}~\bibnamefont {Govenius}},\ }\bibfield  {title} {\bibinfo {title} {Qubit-compatible substrates with superconducting through-silicon vias},\ }\href {https://doi.org/10.1109/TQE.2022.3209881} {\bibfield  {journal} {\bibinfo  {journal} {IEEE Transactions on Quantum Engineering}\ }\textbf {\bibinfo {volume} {3}},\ \bibinfo {pages} {1} (\bibinfo {year} {2022})}\BibitemShut {NoStop}%
\bibitem [{\citenamefont {Schulz}\ \emph {et~al.}(2013)\citenamefont {Schulz}, \citenamefont {Eppelt},\ and\ \citenamefont {Poprawe}}]{Schulz2013}%
  \BibitemOpen
  \bibfield  {author} {\bibinfo {author} {\bibfnamefont {W.}~\bibnamefont {Schulz}}, \bibinfo {author} {\bibfnamefont {U.}~\bibnamefont {Eppelt}},\ and\ \bibinfo {author} {\bibfnamefont {R.}~\bibnamefont {Poprawe}},\ }\bibfield  {title} {\bibinfo {title} {{Review on laser drilling I. Fundamentals, modeling, and simulation}},\ }\href {https://doi.org/10.2351/1.4773837} {\bibfield  {journal} {\bibinfo  {journal} {Journal of Laser Applications}\ }\textbf {\bibinfo {volume} {25}},\ \bibinfo {pages} {012006} (\bibinfo {year} {2013})}\BibitemShut {NoStop}%
\bibitem [{\citenamefont {Jia}\ \emph {et~al.}(2022)\citenamefont {Jia}, \citenamefont {Chen}, \citenamefont {Liu}, \citenamefont {Wang},\ and\ \citenamefont {Duan}}]{Jia2022}%
  \BibitemOpen
  \bibfield  {author} {\bibinfo {author} {\bibfnamefont {X.}~\bibnamefont {Jia}}, \bibinfo {author} {\bibfnamefont {Y.}~\bibnamefont {Chen}}, \bibinfo {author} {\bibfnamefont {L.}~\bibnamefont {Liu}}, \bibinfo {author} {\bibfnamefont {C.}~\bibnamefont {Wang}},\ and\ \bibinfo {author} {\bibfnamefont {J.}~\bibnamefont {Duan}},\ }\bibfield  {title} {\bibinfo {title} {Advances in laser drilling of structural ceramics},\ }\href@noop {} {\bibfield  {journal} {\bibinfo  {journal} {Nanomaterials (Basel)}\ }\textbf {\bibinfo {volume} {12}} (\bibinfo {year} {2022})}\BibitemShut {NoStop}%
\bibitem [{\citenamefont {Muthusubramanian}\ \emph {et~al.}(2024)\citenamefont {Muthusubramanian}, \citenamefont {Finkel}, \citenamefont {Duivestein}, \citenamefont {Zachariadis}, \citenamefont {van~der Meer}, \citenamefont {Veen}, \citenamefont {Beekman}, \citenamefont {Stavenga}, \citenamefont {Bruno},\ and\ \citenamefont {DiCarlo}}]{Muthusubramanian_2024}%
  \BibitemOpen
  \bibfield  {author} {\bibinfo {author} {\bibfnamefont {N.}~\bibnamefont {Muthusubramanian}}, \bibinfo {author} {\bibfnamefont {M.}~\bibnamefont {Finkel}}, \bibinfo {author} {\bibfnamefont {P.}~\bibnamefont {Duivestein}}, \bibinfo {author} {\bibfnamefont {C.}~\bibnamefont {Zachariadis}}, \bibinfo {author} {\bibfnamefont {S.~L.~M.}\ \bibnamefont {van~der Meer}}, \bibinfo {author} {\bibfnamefont {H.~M.}\ \bibnamefont {Veen}}, \bibinfo {author} {\bibfnamefont {M.~W.}\ \bibnamefont {Beekman}}, \bibinfo {author} {\bibfnamefont {T.}~\bibnamefont {Stavenga}}, \bibinfo {author} {\bibfnamefont {A.}~\bibnamefont {Bruno}},\ and\ \bibinfo {author} {\bibfnamefont {L.}~\bibnamefont {DiCarlo}},\ }\bibfield  {title} {\bibinfo {title} {Wafer-scale uniformity of dolan-bridge and bridgeless manhattan-style josephson junctions for superconducting quantum processors},\ }\href {https://doi.org/10.1088/2058-9565/ad199c} {\bibfield  {journal} {\bibinfo  {journal} {Quantum Science and Technology}\ }\textbf {\bibinfo {volume} {9}},\
  \bibinfo {pages} {025006} (\bibinfo {year} {2024})}\BibitemShut {NoStop}%
\bibitem [{\citenamefont {Kreikebaum}\ \emph {et~al.}(2020)\citenamefont {Kreikebaum}, \citenamefont {O’Brien}, \citenamefont {Morvan},\ and\ \citenamefont {Siddiqi}}]{Kreikebaum2020}%
  \BibitemOpen
  \bibfield  {author} {\bibinfo {author} {\bibfnamefont {J.~M.}\ \bibnamefont {Kreikebaum}}, \bibinfo {author} {\bibfnamefont {K.~P.}\ \bibnamefont {O’Brien}}, \bibinfo {author} {\bibfnamefont {A.}~\bibnamefont {Morvan}},\ and\ \bibinfo {author} {\bibfnamefont {I.}~\bibnamefont {Siddiqi}},\ }\bibfield  {title} {\bibinfo {title} {Improving wafer-scale josephson junction resistance variation in superconducting quantum coherent circuits},\ }\href {https://doi.org/10.1088/1361-6668/ab8617} {\bibfield  {journal} {\bibinfo  {journal} {Superconductor Science and Technology}\ }\textbf {\bibinfo {volume} {33}},\ \bibinfo {pages} {06LT02} (\bibinfo {year} {2020})}\BibitemShut {NoStop}%
\bibitem [{\citenamefont {Hertzberg}\ \emph {et~al.}(2020)\citenamefont {Hertzberg}, \citenamefont {Zhang}, \citenamefont {Rosenblatt}, \citenamefont {Magesan}, \citenamefont {Smolin}, \citenamefont {Yau}, \citenamefont {Adiga}, \citenamefont {Sandberg}, \citenamefont {Brink}, \citenamefont {Chow},\ and\ \citenamefont {Orcutt}}]{Hertzberg2020}%
  \BibitemOpen
  \bibfield  {author} {\bibinfo {author} {\bibfnamefont {J.~B.}\ \bibnamefont {Hertzberg}}, \bibinfo {author} {\bibfnamefont {E.~J.}\ \bibnamefont {Zhang}}, \bibinfo {author} {\bibfnamefont {S.}~\bibnamefont {Rosenblatt}}, \bibinfo {author} {\bibfnamefont {E.}~\bibnamefont {Magesan}}, \bibinfo {author} {\bibfnamefont {J.~A.}\ \bibnamefont {Smolin}}, \bibinfo {author} {\bibfnamefont {J.-B.}\ \bibnamefont {Yau}}, \bibinfo {author} {\bibfnamefont {V.~P.}\ \bibnamefont {Adiga}}, \bibinfo {author} {\bibfnamefont {M.}~\bibnamefont {Sandberg}}, \bibinfo {author} {\bibfnamefont {M.}~\bibnamefont {Brink}}, \bibinfo {author} {\bibfnamefont {J.~M.}\ \bibnamefont {Chow}},\ and\ \bibinfo {author} {\bibfnamefont {J.~S.}\ \bibnamefont {Orcutt}},\ }\bibfield  {title} {\bibinfo {title} {{Laser-annealing Josephson junctions for yielding scaled-up superconducting quantum processors}},\ }\href {https://doi.org/10.1038/s41534-021-00464-5} {\bibfield  {journal} {\bibinfo  {journal} {npj Quantum Information}\ }\textbf {\bibinfo
  {volume} {7}},\ \bibinfo {pages} {129} (\bibinfo {year} {2020})}\BibitemShut {NoStop}%
\bibitem [{\citenamefont {Balaji}\ \emph {et~al.}(2024)\citenamefont {Balaji}, \citenamefont {Acharya}, \citenamefont {Armstrong}, \citenamefont {Crawford}, \citenamefont {Danilin}, \citenamefont {Dixon}, \citenamefont {Kennedy}, \citenamefont {Pothuraju}, \citenamefont {Shahbazi},\ and\ \citenamefont {Shelly}}]{Balaji2024}%
  \BibitemOpen
  \bibfield  {author} {\bibinfo {author} {\bibfnamefont {Y.}~\bibnamefont {Balaji}}, \bibinfo {author} {\bibfnamefont {N.}~\bibnamefont {Acharya}}, \bibinfo {author} {\bibfnamefont {R.}~\bibnamefont {Armstrong}}, \bibinfo {author} {\bibfnamefont {K.~G.}\ \bibnamefont {Crawford}}, \bibinfo {author} {\bibfnamefont {S.}~\bibnamefont {Danilin}}, \bibinfo {author} {\bibfnamefont {T.}~\bibnamefont {Dixon}}, \bibinfo {author} {\bibfnamefont {O.~W.}\ \bibnamefont {Kennedy}}, \bibinfo {author} {\bibfnamefont {R.~D.}\ \bibnamefont {Pothuraju}}, \bibinfo {author} {\bibfnamefont {K.}~\bibnamefont {Shahbazi}},\ and\ \bibinfo {author} {\bibfnamefont {C.~D.}\ \bibnamefont {Shelly}},\ }\href@noop {} {\bibinfo {title} {Electron-beam annealing of josephson junctions for frequency tuning of quantum processors}} (\bibinfo {year} {2024}),\ \Eprint {https://arxiv.org/abs/2402.17395} {arXiv:2402.17395 [quant-ph]} \BibitemShut {NoStop}%
\bibitem [{\citenamefont {Pappas}\ \emph {et~al.}(2024)\citenamefont {Pappas}, \citenamefont {Field}, \citenamefont {Kopas}, \citenamefont {Howard}, \citenamefont {Wang}, \citenamefont {Lachman}, \citenamefont {Zhou}, \citenamefont {Oh}, \citenamefont {Yadavalli}, \citenamefont {Sete}, \citenamefont {Bestwick}, \citenamefont {Kramer},\ and\ \citenamefont {Mutus}}]{Pappas2024}%
  \BibitemOpen
  \bibfield  {author} {\bibinfo {author} {\bibfnamefont {D.~P.}\ \bibnamefont {Pappas}}, \bibinfo {author} {\bibfnamefont {M.}~\bibnamefont {Field}}, \bibinfo {author} {\bibfnamefont {C.}~\bibnamefont {Kopas}}, \bibinfo {author} {\bibfnamefont {J.~A.}\ \bibnamefont {Howard}}, \bibinfo {author} {\bibfnamefont {X.}~\bibnamefont {Wang}}, \bibinfo {author} {\bibfnamefont {E.}~\bibnamefont {Lachman}}, \bibinfo {author} {\bibfnamefont {L.}~\bibnamefont {Zhou}}, \bibinfo {author} {\bibfnamefont {J.}~\bibnamefont {Oh}}, \bibinfo {author} {\bibfnamefont {K.}~\bibnamefont {Yadavalli}}, \bibinfo {author} {\bibfnamefont {E.~A.}\ \bibnamefont {Sete}}, \bibinfo {author} {\bibfnamefont {A.}~\bibnamefont {Bestwick}}, \bibinfo {author} {\bibfnamefont {M.~J.}\ \bibnamefont {Kramer}},\ and\ \bibinfo {author} {\bibfnamefont {J.~Y.}\ \bibnamefont {Mutus}},\ }\href@noop {} {\bibinfo {title} {Alternating bias assisted annealing of amorphous oxide tunnel junctions}} (\bibinfo {year} {2024}),\ \Eprint
  {https://arxiv.org/abs/2401.07415} {arXiv:2401.07415 [physics.app-ph]} \BibitemShut {NoStop}%
\bibitem [{\citenamefont {Burnett}\ \emph {et~al.}(2019)\citenamefont {Burnett}, \citenamefont {Bengtsson}, \citenamefont {Scigliuzzo}, \citenamefont {Niepce}, \citenamefont {Kudra}, \citenamefont {Delsing},\ and\ \citenamefont {Bylander}}]{Burnett2019}%
  \BibitemOpen
  \bibfield  {author} {\bibinfo {author} {\bibfnamefont {J.~J.}\ \bibnamefont {Burnett}}, \bibinfo {author} {\bibfnamefont {A.}~\bibnamefont {Bengtsson}}, \bibinfo {author} {\bibfnamefont {M.}~\bibnamefont {Scigliuzzo}}, \bibinfo {author} {\bibfnamefont {D.}~\bibnamefont {Niepce}}, \bibinfo {author} {\bibfnamefont {M.}~\bibnamefont {Kudra}}, \bibinfo {author} {\bibfnamefont {P.}~\bibnamefont {Delsing}},\ and\ \bibinfo {author} {\bibfnamefont {J.}~\bibnamefont {Bylander}},\ }\bibfield  {title} {\bibinfo {title} {Decoherence benchmarking of superconducting qubits},\ }\href {https://doi.org/10.1038/s41534-019-0168-5} {\bibfield  {journal} {\bibinfo  {journal} {npj Quantum Information}\ }\textbf {\bibinfo {volume} {5}},\ \bibinfo {pages} {54} (\bibinfo {year} {2019})}\BibitemShut {NoStop}%
\bibitem [{OQC(2023)}]{OQC_Lucy}%
  \BibitemOpen
  \href {https://aws.amazon.com/braket/quantum-computers/oqc/} {\bibinfo {title} {{Oxford Quantum Circuits}}},\ \bibinfo {howpublished} {\url{https://aws.amazon.com/braket/quantum-computers/oqc/}} (\bibinfo {year} {2023})\BibitemShut {NoStop}%
\bibitem [{\citenamefont {Dolan}(1977)}]{Dolan1977}%
  \BibitemOpen
  \bibfield  {author} {\bibinfo {author} {\bibfnamefont {G.~J.}\ \bibnamefont {Dolan}},\ }\bibfield  {title} {\bibinfo {title} {Offset masks for lift‐off photoprocessing},\ }\href {https://doi.org/10.1063/1.89690} {\bibfield  {journal} {\bibinfo  {journal} {Applied Physics Letters}\ }\textbf {\bibinfo {volume} {31}},\ \bibinfo {pages} {337} (\bibinfo {year} {1977})}\BibitemShut {NoStop}%
\bibitem [{\citenamefont {Ahmad}\ \emph {et~al.}(2014)\citenamefont {Ahmad}, \citenamefont {Haring}, \citenamefont {Reinholz},\ and\ \citenamefont {Schneck}}]{Ahmad2014}%
  \BibitemOpen
  \bibfield  {author} {\bibinfo {author} {\bibfnamefont {S.~S.}\ \bibnamefont {Ahmad}}, \bibinfo {author} {\bibfnamefont {F.}~\bibnamefont {Haring}}, \bibinfo {author} {\bibfnamefont {A.}~\bibnamefont {Reinholz}},\ and\ \bibinfo {author} {\bibfnamefont {N.}~\bibnamefont {Schneck}},\ }\bibfield  {title} {\bibinfo {title} {Through-sapphire via filling process development for backside interconnect and chip stacking and front-to-back daisy chain chip},\ }\href {https://doi.org/10.4071/2014DPC-wp11} {\bibfield  {journal} {\bibinfo  {journal} {IMAPSource Proceedings}\ }\textbf {\bibinfo {volume} {2014}},\ \bibinfo {pages} {1422} (\bibinfo {year} {2014})}\BibitemShut {NoStop}%
\bibitem [{\citenamefont {Lisenfeld}\ \emph {et~al.}(2019)\citenamefont {Lisenfeld}, \citenamefont {Bilmes}, \citenamefont {Megrant}, \citenamefont {Barends}, \citenamefont {Kelly}, \citenamefont {Klimov}, \citenamefont {Weiss}, \citenamefont {Martinis},\ and\ \citenamefont {Ustinov}}]{Lisenfeld2019}%
  \BibitemOpen
  \bibfield  {author} {\bibinfo {author} {\bibfnamefont {J.}~\bibnamefont {Lisenfeld}}, \bibinfo {author} {\bibfnamefont {A.}~\bibnamefont {Bilmes}}, \bibinfo {author} {\bibfnamefont {A.}~\bibnamefont {Megrant}}, \bibinfo {author} {\bibfnamefont {R.}~\bibnamefont {Barends}}, \bibinfo {author} {\bibfnamefont {J.}~\bibnamefont {Kelly}}, \bibinfo {author} {\bibfnamefont {P.}~\bibnamefont {Klimov}}, \bibinfo {author} {\bibfnamefont {G.}~\bibnamefont {Weiss}}, \bibinfo {author} {\bibfnamefont {J.~M.}\ \bibnamefont {Martinis}},\ and\ \bibinfo {author} {\bibfnamefont {A.~V.}\ \bibnamefont {Ustinov}},\ }\bibfield  {title} {\bibinfo {title} {Electric field spectroscopy of material defects in transmon qubits},\ }\href {https://doi.org/10.1038/s41534-019-0224-1} {\bibfield  {journal} {\bibinfo  {journal} {npj Quantum Information}\ }\textbf {\bibinfo {volume} {5}},\ \bibinfo {pages} {105} (\bibinfo {year} {2019})}\BibitemShut {NoStop}%
\end{thebibliography}%

\newpage

\appendix 

\end{document}